\documentclass[11pt,amsmath,amssymb]{revtex4}
\usepackage{amsfonts,epsfig,psfrag}

\newcommand {\jn}     {\uparrow}
\newcommand {\ur}     {\downarrow}
\newcommand {\kB}     {\kappa_{\!_B}}
\newcommand {\uB}     {\mu_{_B}}

\begin{document}

\title{Phase diagram of a coupled tetrahedral Heisenberg model}

\author{Onofre Rojas and F. C. Alcaraz}

\affiliation{ Universidade de S\~ao Paulo, Instituto de F\'{\i}sica de S\~ao Carlos, \\ C.P. 369, 13560-590, S\~ao Carlos, SP Brazil}

\date{\today}

\begin{abstract}
The phase diagram of a coupled tetrahedral Heisenberg model is obtained. The 
quantum chain has a local gauge symmetry and its eigenspectrum is obtained 
by the  composition of the eigenspectra
of  spin-$\frac{1}{2}$ XXZ chains with
arbitrary distribution of spin-$\frac{3}{2}$ impurities. The phase diagram is 
quite rich with an infinite number of phases with ferromagnetic,
antiferromagnetic or ferrimagnetic order. In some cases the ground state  and 
the low lying eigenlevels of 
the model can be exactly calculated since they  coincide with the 
eigenlevels of the exactly integrable XXZ chain. The thermodynamical 
properties of the model at low temperatures is also studied through finite-size
analysis.  
\end{abstract}

\maketitle

\section{Introduction}

\sloppy
Recently an extensive amount of experimental and theoretical investigations 
on a large variety of quasi-one-dimensional mixed spin compounds have 
been reported in the literature \cite{sakaiprb98}-\cite{koga_cia}.
One of the main reason for the  interest 
on  these physical systems is due to the richness of their magnetic 
properties, exhibiting 
antiferromagnetism, ferromagnetism, ferrimagnetism, magnetization 
plateaux, etc. 

 The simplest model on this class is the mixed spin-($S,S'$) Heisenberg model
 where we have two types of antiferromagnetically exchange-coupled 
 spins $S$ and $S'$ located at alternated sites of a chain. According to the 
 Lieb-Mattis theorem \cite{liebmattis} in the cases where $S \neq S'$ the model
 exhibits ferrimagnetic behavior, having a $(S-S')\frac{L}{2}$-degenerated 
 ground state, where $L$ is the chain length.   
 Extension of these models to quantum chains with two or more different 
 quantum spins in the elementary cells increases considerably the number of
 quite interesting models, since the topology 
 besides allowing  a mechanism of frustration of spins also  determines  
 the ordering of the ground 
 state, and the nature of the low-lying excitations.  Several of these 
 mixed-spin compounds exhibit local gauge invariance rendering  a
 great amount of simplification  on the analysis of these quantum chains 
 through finite-size studies.
 In \cite{alcaraz_malvezzi} conformal invariance was exploited to obtain the
 critical properties of the simplest mixed-spin compound where we have at the 
 odd sites a spin $\frac{1}{2}$ connected with two disconnected spin
 $\frac{1}{2}$  at even sites. In \cite{niggemannG,niggemannT}, by
  introducing interactions among the spins at the even sites it was shown that
  phase transitions occur as the strength of the interactions at the even sites
  changes. Another variant of these alternating compound spin chains is the 
  orthogonal-dimer spin chain investigated in
  \cite{ivanov_richter}, who is also known as 
  the frustrated dimer-plaquete chain.
  All these quantum chains due to its gauge symmetry have  a phase transition 
  to a phase whose ground state in the thermodynamic limit is fragmented, 
  i. e., it is formed by the composition of ground states  of finite quantum
  chains with open ends.

 Motivated by these results we introduce and study in this paper a 
 mixed-compound quantum chain that, as the previous models, exhibits a 
 local gauge symmetry, but distinctly does not exhibit the fragmentation 
 properties. These models describe the dynamics of 
  spin-$\frac{1}{2}$  Pauli matrices  located at the vertices of  
  tetrahedrals and 
  interacting as in Fig. 1. 
 As we shall see this quantum chain reveals a quite rich phase diagram with 
 ferromagnetic, antiferromagnetic, ferrimagnetic massive phases as well 
 massless disordered phases. These phases are connected  through 
 continuous and discontinuous phase
 transitions, as we change the coupling constants. The number of phase 
 transitions can be finite or infinite, depending on the region of 
 the phase diagram. 

 Another interesting point about our model concerns the question of exact 
 integrability of quantum chains. We show that part of the eigenenergies of 
 our model  are just the eigenenergies of the exact integrable 
 spin-$\frac{1}{2}$ XXZ chain, whose eigenenergies are exactly known 
 through the Bethe ansatz \cite{yangyang}. Depending on the exchange 
 couplings  these eigenenergies become the low lying ones, including the 
 ground state energy. Consequently our model provides a clear example of a
 non-integrable model that contains part of its eigenspectrum exactly 
 integrable. An interesting feature of  our tetrahedral spin chain 
 is the appearance of  phases 
 with a non-zero entropy per site in the thermodynamical limit, similarly 
 as happen with the residual entropy of ice \cite{pauling}.

\begin{figure}[!ht]
\begin{center}
 \psfrag{t}[]{$\pmb\tau$}
 \psfrag{s}[]{$\pmb\sigma$}
\includegraphics[width=7.5cm,height=2.7cm]{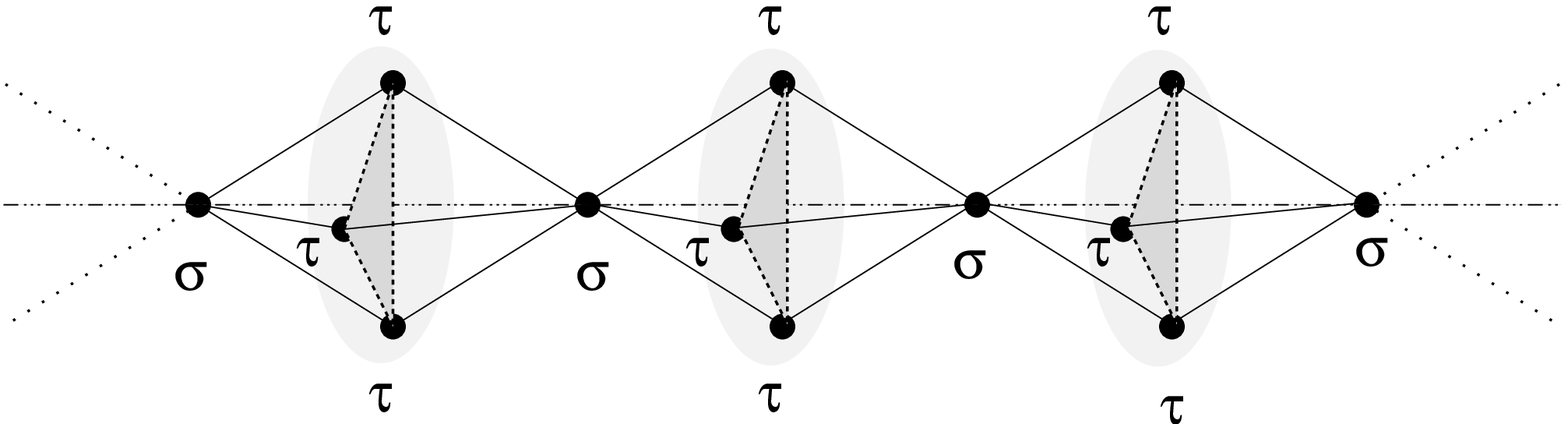}
\caption[fig1]{One dimensional array of tetrahedral spin cluster, where $\pmb\sigma$ and $\pmb\tau$ are spin-$\frac{1}{2}$  matrices, dashed lines  correspond to the exchange interaction $J$ in the triangle configuration and  the solid lines represent the 
anisotropic interaction  among the spins $\pmb\sigma$  and  $\pmb\tau$ spins,  with  anisotropy  $\Delta$.}
\label{fig1}
\end{center} 
\end{figure}

The paper is organized as follows. In \S2 we present our model and by exploring
its gauge symmetry we show that its eigenspectrum is given by the composition
of the eigenspectrum of spin-$\frac{1}{2}$ XXZ chains with 
spin-$\frac{3}{2}$ impurities. In \S3 we study the spectral properties 
of the spin-$\frac{1}{2}$ XXZ chain in the presence of spin-$\frac{3}{2}$
impurities.  In \S4 using numerical calculations and the results of previous 
sections we obtain the phase diagram of our  tetrahedral spin chain introduced 
in this paper. In \S5 we study the thermodynamical properties of the 
tetrahedral spin chain by evaluating its entropy, specific heat and the 
magnetic susceptibility at zero magnetic field. Finally in \S6 we present 
our general conclusions.

\section{The Model}
The model we introduce in this paper describes the dynamics of the spin-$\frac{1}{2}$ spins located in a one dimensional array of tetrahedral spin clusters as shown in Fig. \ref{fig1}. The spin operators attached to the basis and to the top of these tetrahedral are denoted by $\pmb\sigma_i$ ($i=1,2,\dots,\frac{L}{2}$) and $\pmb\tau^a_i$, $\pmb\tau^b_i$ and $\pmb\tau^c_i$ ($i=1,2,\dots,\frac{L}{2}$), respectively. The interactions are among  nearest  neighbors  in this geometrical structure (see Fig. \ref{fig1}),  and the Hamiltonian in a $\frac{L}{2}$ tetrahedral spin chain ($L$ even) is given by

\begin{widetext}
\begin{equation}\label{hamiltonian}
{\bf H}=-\sum_{i=1}^{L/2}\Big(({\pmb\sigma}_i,{\pmb{\mathcal T}}_i)_{\Delta}+({\pmb\sigma}_{i+1},{\pmb{\mathcal T}}_i)_{\Delta}-J\big({\pmb \tau}_i^a{\pmb \tau}_i^b+{\pmb \tau}^b_i{\pmb \tau}^c_i+{\pmb \tau}^c_i{\pmb \tau}^a_i\big)\Big)
\end{equation}
\end{widetext}
where
\begin{equation}
{\pmb{\mathcal T}}_i =({\mathcal T}_i^x,{\mathcal T}_i^y,{\mathcal T}_i^z)=\pmb\tau_i^a+\pmb\tau_i^b+\pmb\tau_i^c,
\end{equation}

\begin{equation}
({\pmb\sigma}_i,{\pmb{\mathcal T}}_j)_{\Delta}=\sigma_i^x{\mathcal T}_j^x+\sigma_i^y{\mathcal T}_j^y+\Delta\sigma_i^z{\mathcal T}_j^z,
\end{equation}
$i=1,2,\dots,\frac{L}{2}$, and we have imposed periodic boundary conditions. The spin operators ($\pmb\tau_i^a,\pmb\tau_i^b,\pmb\tau_i^c$) have isotropic interactions among themselves (coupling $J$) and XXZ-like interaction (anisotropy $\Delta$) with the spin operators $\pmb\sigma$. The particular case where $\Delta=-1$ coincides with
 a special point of an alternated tetrahedral spin chain introduced in \cite{mambrini}.  The competitions between these two type of interactions will produce a rich phase diagram.  Clearly if $J$ is negative the $\pmb{\pmb\tau}$ operators prefer to align forming a spin-$\frac{3}{2}$ operator and we should expect an effective mixed anisotropic spin-($\frac{1}{2},\frac{3}{2}$) Heisenberg chain.  However when $J>0$ we necessarily have frustrations among the  $\pmb\tau$ interactions in the basis of the tetrahedron, but the amount of these frustrations will depend upon the relative strength of the anisotropic interaction  with the operators $\pmb\sigma$.  For $|J|\lesssim 1$ we should expect that the XXZ interactions being stronger will induce, as in the case $J<0$,  an effective mixed spin-($\frac{1}{2},\frac{3}{2}$) Heisenberg chain. On the other hand for $J\gtrsim 1$ the interactions among the $\pmb\tau$ operators, defined on the basis of the tetrahedron, are more important and in order to minimize their frustration they cluster into   spin $s=\frac{1}{2}$ operators and the Hamiltonian should behave effectively as an non-mixed and standard anisotropic spin-$\frac{1}{2}$ Heisenberg chain or XXZ quantum chain. Since the isotropic mixed spin-($\frac{3}{2},\frac{1}{2}$) Heisenberg chain exhibits ferrimagnetic behavior in contrast with the standard antiferromagnetic behavior of the spin-$\frac{1}{2}$ XXZ   chain, we should expect phase transitions lines separating the above regimes.

The Hamiltonian \eqref{hamiltonian} exhibit a local gauge symmetry that will render a great simplification in our physical and numerical analysis of the model.  It is simple to verify that the local operators 

\begin{equation}\label{gauge}
{\bf G}_i={\pmb{\mathcal T}}_i^2=\big({\pmb \tau}_i^a+{\pmb \tau}_i^b+{\pmb \tau}_i^c\big)^2,\qquad i=1,2,\dots,\frac{L}{2},
\end{equation}
besides commuting among themselves commute  with the Hamiltonian \eqref{hamiltonian}. Consequently in the basis where all the $\{{\bf G}_i\}$ are diagonal the Hilbert space associated to \eqref{hamiltonian} can be separated into block disjoint sectors labeled by the eigenvalues $\{g_i\}$ of the operators $\{{\bf G}_i\}$  ($i=1,2,\dots,\frac{L}{2}$). Since $\pmb{\mathcal T}_i$ is formed by the addition of three spin-$\frac{1}{2}$ angular momentum operators its eigenvalues are $g_i=S(S+1)$, with $S=\frac{3}{2},\frac{1}{2}$. The eigenvectors corresponding to $g_i=\frac{15}{4}$, in the ${\mathcal T}^z$-basis can be written as the quadruplet
\begin{subequations}\label{eigenvec3}
\begin{align}\nonumber\label{quadroplet}
\big|\frac{3}{2},\frac{3}{2}\big\rangle&\;=\;|{+\!+\!+}\rangle\\\nonumber
\big|\frac{3}{2},\frac{1}{2}\big\rangle&\;=\;\frac{1}{\sqrt{3}}\Big(|{+\!+\!-}\rangle\;+\;|{+\!-\!+}\rangle\;+\;|{-\!+\!+}\rangle\Big)\\\nonumber
\big|\frac{3}{2},\frac{-1}{2}\big\rangle&\;=\;\frac{1}{\sqrt{3}}\Big(|{-\!-\!+}\rangle\;+\;|{-\!+\!-}\rangle\;+\;|{+\!-\!-}\rangle\Big)\\
\big|\frac{3}{2},\frac{-3}{2}\big\rangle&\;=\;|{-\!-\!-}\rangle,
\end{align}
while the eigenvectors corresponding to  $g_i=\frac{3}{4}$ can be written as the quadruplet
\begin{align}\nonumber\label{doublet_a}
\big|\frac{1}{2},\frac{1}{2}\big\rangle&\;=\;\frac{1}{\sqrt{2}}\Big(|{-\!+\!+}\rangle\;-\;|{+\!+\!-}\rangle\Big)\\
\big|\frac{1}{2},\frac{-1}{2}\big\rangle&\;=\;\frac{1}{\sqrt{2}}\Big(|{+\!-\!-}\rangle\;-\;|{-\!-\!+}\rangle\Big)
\end{align}
and
\begin{align}\nonumber\label{doublet_b}
\big|\frac{1}{2},\frac{1}{2}\big\rangle^{\prime}&=\frac{1}{\sqrt{6}}\Big(|{-\!+\!+}\rangle\;-2\;|{+\!-\!+}\rangle\;+\;|{+\!+\!-}\rangle\Big)\\
\big|\frac{1}{2},\frac{-1}{2}\big\rangle^{\prime}&=\frac{1}{\sqrt{6}}\Big(|{+\!-\!-}\rangle\;-2\;|{-\!+\!-}\rangle\;+\;|{-\!-\!+}\rangle\Big).
\end{align}
\end{subequations}

As a consequence of the local gauge invariance the action of the Hamiltonian on the basis vectors only connects vectors with the same eigenvalue $\{g_i\}$. Moreover in the case where $g_i=\frac{3}{4}$ the first two eigenvectors given in \eqref{doublet_a} are not coupled through the interactions with the last ones, given in \eqref{doublet_b}.  This means that in fact the gauge choice $g_i=\frac{1}{2}$, for a given site $i$, can be chosen on two independent ways. 

Therefore using the basis \eqref{eigenvec3} the Hilbert space associated to \eqref{hamiltonian} are separated into sectors labeled by $\{g_1,g_2,\dots,g_{\frac{L}{2}}\}$, where $g_i=\frac{15}{4},\frac{3}{4},\frac{3}{4}$.  For a given choice $\{g_i\}$ 
($i=1,\ldots,\frac{L}{2}$) we verify that in this restricted subspace the effective Hamiltonian reduces to  a general mixed anisotropic Heisenberg chain

\begin{widetext}

\begin{eqnarray}\label{Hamt_g}
{\bf H}_{\{g_1,\dots,g_{\frac{L}{2}}\}}&=&-\sum_{i=1}^{L/2}
\Big(\sigma_{2i-1}^x S_{2i}^x+S_{2i}^x\sigma_{2i+1}^x +
\sigma_{2i-1}^y S_{2i}^y+ S_{2i}^y\sigma_{2i+1}^y + 
\Delta(\sigma_{2i-1}^z S_{2i}^z + S_{2i}^z\sigma_{2i+1}^z) \nonumber
\\ &&+ \frac{J}{2}\sum_{i=1}^{L/2}g_i-J\frac{9}{16}L,
\end{eqnarray}

\end{widetext}
where at the odd sites we have spin-$\frac{1}{2}$ operators $\pmb\sigma_i$, and at the even sites the spin operators ${\bf S}_i$ are  spin-$\frac{3}{2}$ or spin-$\frac{1}{2}$ Pauli matrices depending if $g_i=\frac{15}{4}$ or $g_i=\frac{3}{4}$, respectively.  Consequently a complete understanding of our model will demand the  understanding of the infinite set of general mixed spin quantum chain \eqref{Hamt_g}. Particularly in order to calculate the phase diagram we should know, as a function of $\Delta$ and $J$,   among all these $3^\frac{L}{2}$ mixed Heisenberg chains, which one  provides the global ground state of the model.

Some special gauge choices, according to our previous qualitative analysis are natural candidates for  hosting the ground state in regions of the phase diagram ($J,\Delta$).  The gauge $g_1=g_2=\dots=g_{\frac{L}{2}}=\frac{15}{4}$ give us the anisotropic mixed spin-($\frac{1}{2},\frac{3}{2}$) Heisenberg chain that at $\Delta=-1$, due to the Lieb Mattis theorem \cite{liebmattis} exhibits a $\frac{L}{2}$-fold degenerated ferrimagnetic ground state.  Certainly for large and positive values of $J$ this sector   contains the ground state.

The $2^{\frac{L}{2}}$ distinct choices where $g_1=g_2=\dots=g_{\frac{L}{2}}=\frac{3}{4}$, correspond to degenerated eigensectors with the same eigenspectra as the standard and non-mixed spin-$\frac{1}{2}$ XXZ chain, whose exact solution is known \cite{yangyang} through the Bethe ansatz.

It is interesting  to stress here that although our coupled tetrahedral spin chain \eqref{hamiltonian} is not exact integrable it contains in its eigenspectrum an infinite set of exactly known eigenvalues.  Moreover we should expect that this set of eigenenergies in special regions of the phase diagram ($J,\Delta$) will contain the ground state energy and low lying excitations of the model \eqref{hamiltonian}.  In these regions the ground state of the model will be a $2^{\frac{L}{2}}$-degenerated ground state exactly known through the Bethe ansatz.

Starting from the gauge $g_i=\frac{3}{4}$ ($i=1,2,\dots,\frac{L}{2}$), we may  produce the other gauges by choosing some of $g_i=\frac{15}{4}$. This gives an effective anisotropic spin-$\frac{1}{2}$ Heisenberg chain with magnetic spin-$\frac{3}{2}$ impurities, at the sites  positions $i$ where $g_i=\frac{15}{4}$.

Differently from other quantum chains possessing local gauge symmetry\cite{alcaraz_malvezzi,niggemannG,niggemannT,ivanov_richter} our Hamiltonian \eqref{hamiltonian} does not  produce a fragmentation of the lattice.  The fragmentation happens in those models  \cite{alcaraz_malvezzi,niggemannG,niggemannT,ivanov_richter}, because the effect of gauge fixing is to produce non-magnetic (spin zero) impurities. 

As we can see from \eqref{Hamt_g} the coupling $J$ plays the role of a chemical potential since it gives the additional energy required to insert a spin-$\frac{3}{2}$ impurity into the system.  Before calculating the phase diagram of the Hamiltonian \eqref{Hamt_g} we need to calculate, as a function of the impurity number, the ground state energy of the anisotropic spin-$\frac{1}{2}$ Heisenberg chain with spin-$\frac{3}{2}$ impurities.
 
\section{The Heisenberg chain with impurities}

The Heisenberg chain with impurities were studied in the case of periodic array of impurities by Fukui and Kawakami \cite{fukui_kawakami}.  Their results reveal that whenever the two types of spins defining the chain are odd-half integers the system is gapless.  In the present case, in order to calculate the phase diagram of \eqref{hamiltonian}, we need to know the lowest eigenenergy among all the ground-state energies of the spin-$\frac{1}{2}$ Hamiltonian with a given number of spin-$\frac{3}{2}$ impurities located at the even sites. In general these quantum chains on a periodic lattice are not translational invariant.  For example for $L=20$ the configurations of impurities, apart from trivial rotations by two sites, are given by  the configurations  $C_0,C_1,\dots,C_4$ in Fig. \ref{fig_2S}, and in the case of $L=16$ the configurations with $n=4$ impurities are given by $C_0,C_1,\dots,C_7$ in Fig. \ref{fig_4S}.  We calculate, using the Lanczos method the lowest eigenenergy of these Hamiltonians with arbitrary 
distribution of impurities, for lattice sizes up to $L=20$ or up to $L=16$
depending if the anisotropy $\Delta \le -1$ or  
 $-1\le\Delta\le 1$, respectively. In the case of low density of 
 impurities we are able to extend our calculations up to $L=24$.
 
\begin{widetext}

\begin{figure}[ht]
\begin{center}
\psfrag{s}{$\sigma$}
 \psfrag{S}{$S$}
\psfrag{c0}[]{$C_0$}
 \psfrag{c1}[]{$C_1$}
 \psfrag{c2}[]{$C_2$}
 \psfrag{c3}[]{$C_3$}
 \psfrag{c4}[]{$C_4$}
\includegraphics[width=12cm,height=2cm]{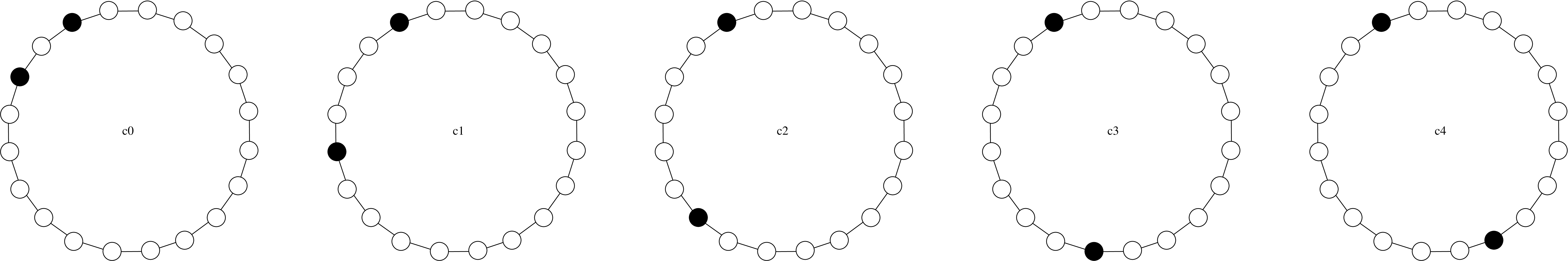}
\end{center}
\caption[fig_2S]{The configurations of two impurities of the chain with $L=20$ sites.}
\label{fig_2S}
\end{figure}

\begin{figure}[ht]
\begin{center}
\psfrag{s}{$\sigma$}
 \psfrag{S}{$S$}
 \psfrag{c0}{$C_0$}
 \psfrag{c1}{$C_1$}
 \psfrag{c2}{$C_2$}
 \psfrag{c3}{$C_3$}
 \psfrag{c4}{$C_4$}
 \psfrag{c5}{$C_5$}
 \psfrag{c6}{$C_6$}
 \psfrag{c7}{$C_7$}
\includegraphics[width=10cm,height=4cm]{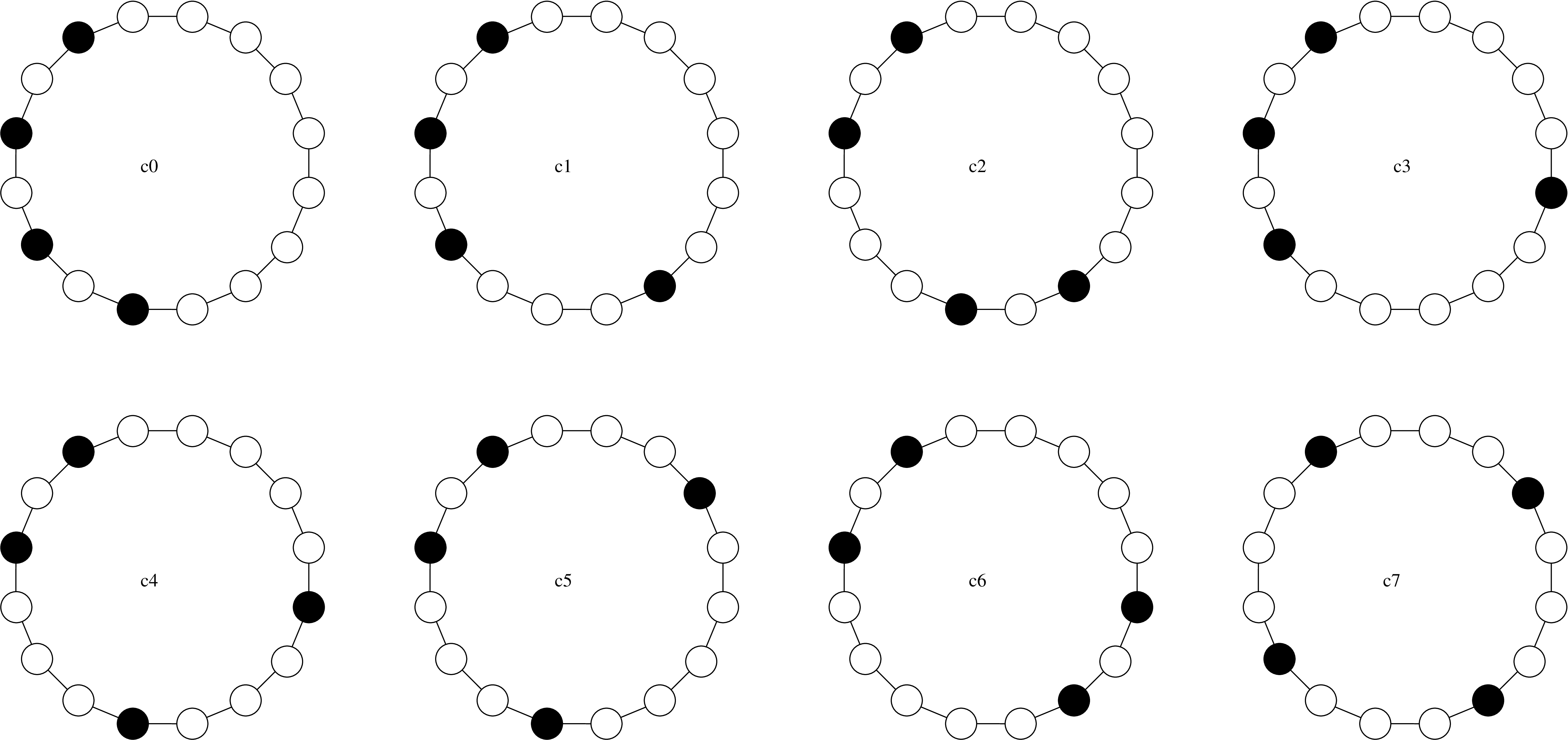}
\caption[fig_4S]{The configurations of 4 impurities of the chain with $L=16$ sites.}
\label{fig_4S}
\end{center}
\end{figure}

\end{widetext}

The  ground-state energy of the Hamiltonians with arbitrary  distributions of $n$ impurities ($0\leqslant n \leqslant \frac{L}{2}$) is non degenerated  for $|\Delta|<1$, two-fold degenerated for $|\Delta|>1$ and at $\Delta=-1$,  due to the Lieb-Mattis theorem \cite{liebmattis},  the ground state is $(2n+1)$-fold degenerated.

\begin{figure}[ht]
\psfrag{DE}[b][][0.9]{$ E_{C_i}-E_{C_0}$}
 \psfrag{D}{\small$\Delta$}
\psfrag{a}{(a)}
\psfrag{b}{(b)}
\psfrag{Ec1------Ec0}[][]{\scriptsize$E_{ C_1}$-$E_{C_0}$}
\psfrag{Ec2-Ec0}{\scriptsize$E_{C_2}$-$E_{C_0}$}
\psfrag{Ec3-Ec0}{\scriptsize$E_{C_3}$-$E_{C_0}$}
\psfrag{Ec4-Ec0}{\scriptsize$E_{C_4}$-$E_{C_0}$}
\psfrag{Ec5-Ec0}{\scriptsize$E_{ C_5}$-$E_{C_0}$}
\psfrag{Ec6-Ec0}{\scriptsize$E_{C_6}$-$E_{C_0}$}
\psfrag{Ec7-Ec0}{\scriptsize$E_{C_7}$-$E_{C_0}$}

\includegraphics[width=6cm,height=7.7cm,angle=-90]{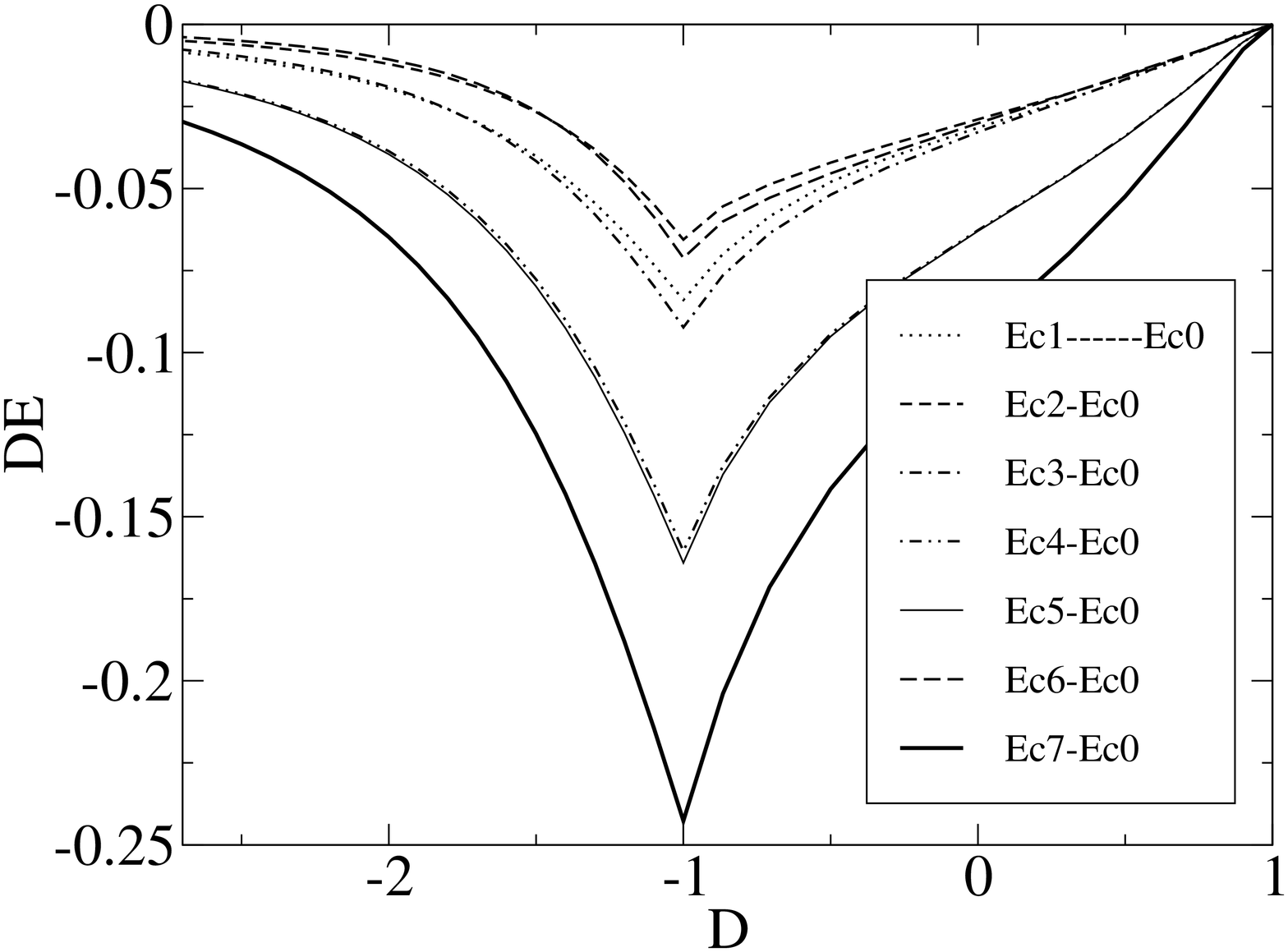}
\caption[fig_behav_2]{The lowest energy of the Hamiltonian with $L=16$ sites and impurities at the 
configurations  $C_1$-$C_7$ of Fig. \ref{fig_4S}. The energies are measured with respect to the 
lowest energy $E_{C_0}$ of the Hamiltonian with the impurity configuration $C_0$. }
\label{fig_behav_2}
\end{figure}

In order to compare the different ground-state energies for the Hamiltonians with distinct impurity configurations we plot in Fig. \ref{fig_behav_2}, as a function of $\Delta$,  the energy differences $E_{C_i}-E_{C_0}$ ($i=1,\ldots,7$) between the ground-state energies of the Hamiltonians with the impurity configurations of Fig. \ref{fig_4S}, for the chain will $L=16$ sites.  We see from this figure that for all values of $\Delta$ the ground-state energy $E_{C_7}$, corresponding to the homogeneous distribution of impurities, has the lowest value, followed by the energies  $E_{C_6}$ and  $E_{C_2}$. The energy $E_{C_0}$, that corresponds to the configuration where the impurities are at the closest positions is the higher one.  This behavior repeats for arbitrary lattice sizes and number of impurities, showing that due to quantum fluctuations the impurities repeal each other favoring the configuration where they are homogeneously distributed.  Our numerical results also show that the minimum energy among these configurations $E_L^0(n,\Delta,0)$ decreases with the total number of impurities, which can be understood since the interaction between two spins $\frac{1}{2}$ and  $\frac{3}{2}$ give lower contributions to the energy than the interactions between two spins $\frac{1}{2}$.

The   lowest eigenenergy $E_L^0(n,\Delta,J)$ of our tetrahedral Hamiltonian \eqref{hamiltonian} in the gauge sector with $n$ values $g_i=\frac{15}{4}$, from \eqref{Hamt_g}, is then given by
\begin{align}\label{E_finito}
\frac{E_L^0(n,\Delta,J)}{L}=\frac{E_L^0(n,\Delta,0)}{L}-\frac{3}{4}(\frac{1}{2}-\frac{2n}{L})J.
\end{align} 

\begin{figure}[ht]
\begin{center}
\psfrag{E}[b][][0.9]{\small$E_L(n,-1,J)/L$}
\psfrag{J}[]{$J$}
\psfrag{L=12}{$L=12$}
\psfrag{L=18}{$L=18$}
\psfrag{L=20}{$L=20$}
\includegraphics[width=5.7cm,height=7.5cm,angle=-90]{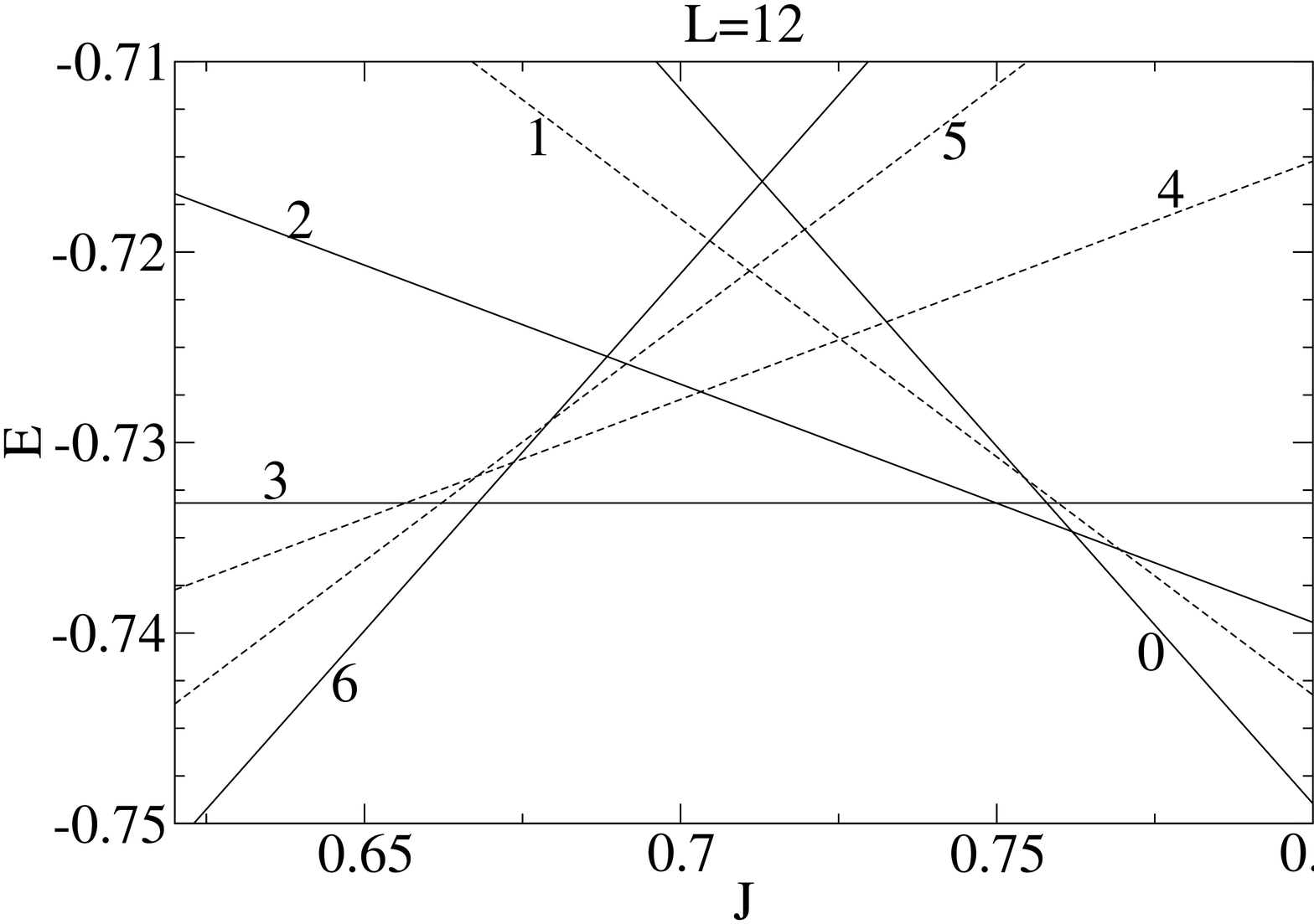}
\includegraphics[width=5.7cm,height=7.5cm,angle=-90]{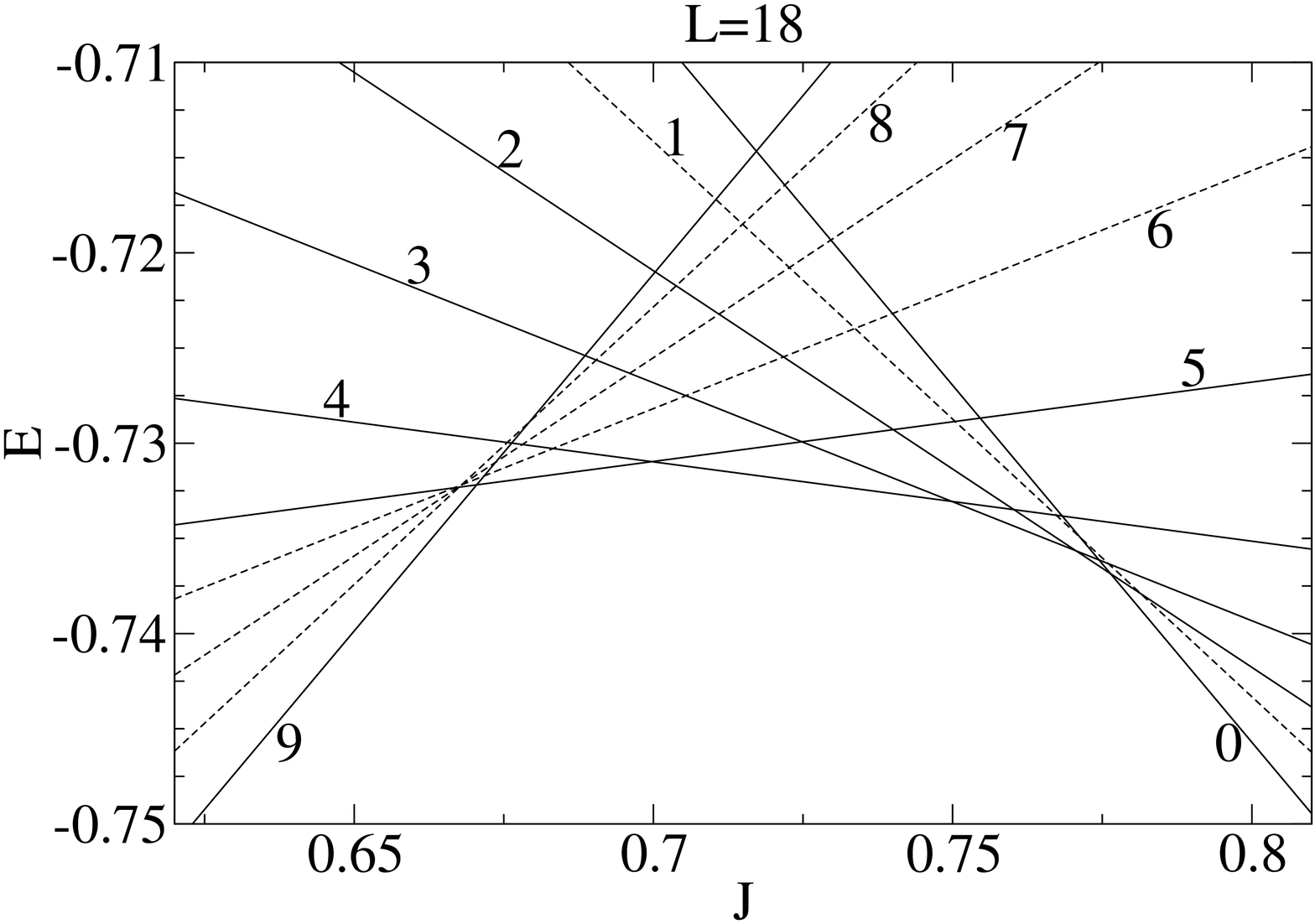}
\includegraphics[width=5.7cm,height=7.5cm,angle=-90]{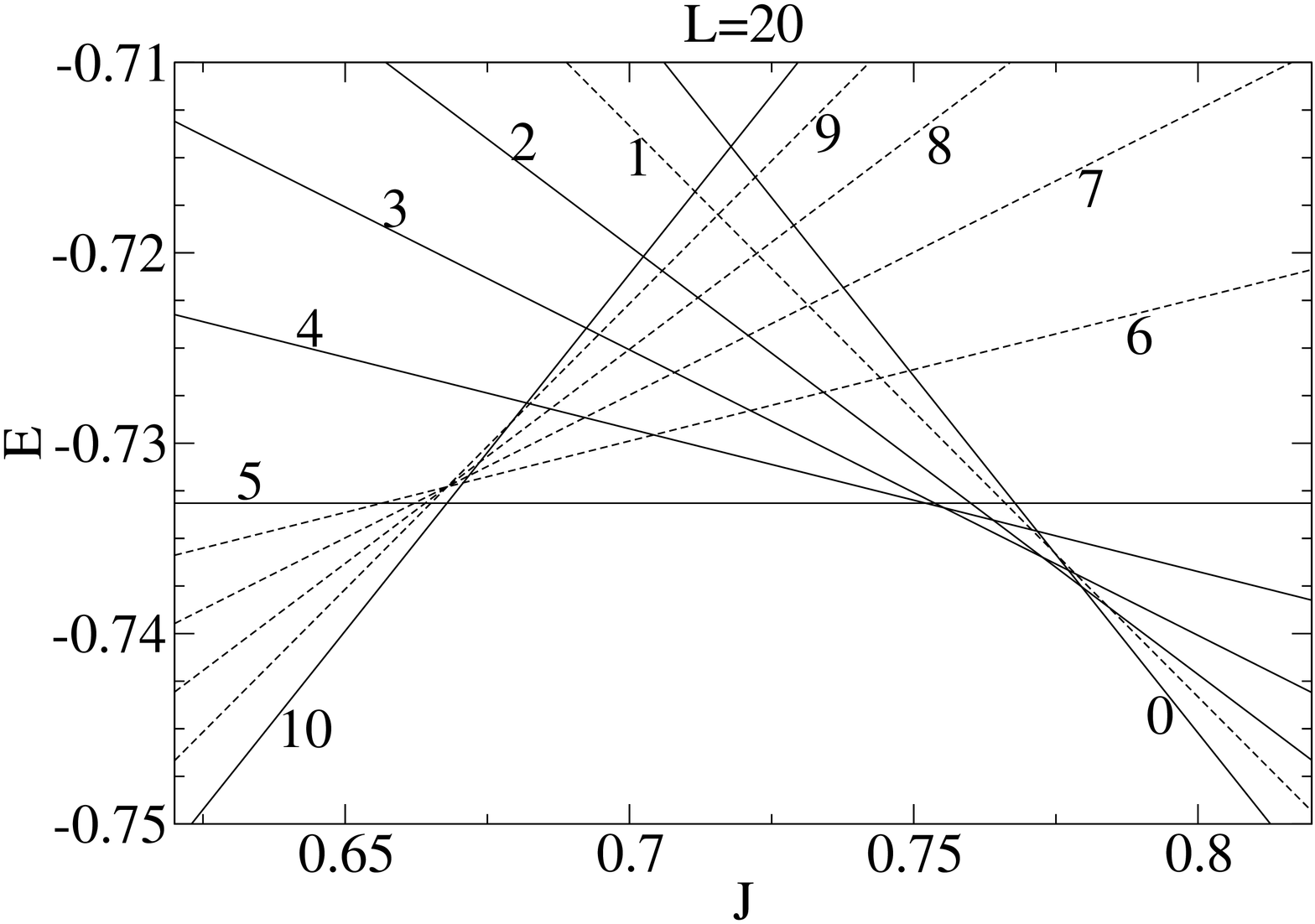}
\caption[fig_diag]{The lowest eigenenergies per site $E_L(n,J)/L$, as a function of $J$, for all the possible 
gauge choices of the finite-size chain with $L=16, 18$ and $20$ with $\Delta =
-1$.  The values of $n$ are shown  in the figures.}
\label{fig_diag}
\end{center}
\end{figure}

From  our previous discussions, these energies have a degeneracy $2^{\frac{L}{2}-n}$ for $|\Delta|<1$, $2\cdot 2^{\frac{L}{2}-n}$ for $\Delta<-1$, $2\binom{L/2}{n} 2^{\frac{L}{2}-n}$ for $\Delta>1$, $(L+1+2n)\binom{L/2}{n}$ for $\Delta=1$ and $(2n+1)\cdot 2^{\frac{L}{2}-n}$ at $\Delta=-1$.  The ground-state energy of our tetrahedral spin chain \eqref{hamiltonian}, for a given value of $J$ and $\Delta$, will be given by the lowest value of \eqref{E_finito}.  In Fig. \ref{fig_diag} we show as a function of $J$ and at $\Delta=-1$ all the lowest eigenenergies $E_L^0(n,-1,J)/L$\; $n=0, 1,2,\dots,\frac{L}{2}$ for the lattice sizes $L=12$, 18 and 20. They are given by straight lines whose slope values depend upon the number $n$, labelling the gauge, and their intercept points with the vertical axis 
($E_L^0(n,-1,0)/L$) increase  as  the number of impurities decreases.  
Our numerical analysis shows that these plots depend strongly on the value of $\Delta$.

We see from Fig. \ref{fig_diag}, in agreement with our earlier physical 
arguments for $\Delta =-1$, that for $J_{\frac{1}{2}}(L,-1)<J<
J_{\frac{1}{4}}(L,-1)$, 
with $J_{\frac{1}{2}}(L,-1)\rightarrow-\infty$ and 
$J_{\frac{1}{4}}(L,-1)\approx 0.667$, the ground state energy is $E_L^0(L/2,-1,J)$ and belongs to the gauge sector $g_1=g_2=\dots=g_{\frac{L}{2}}=\frac{15}{4}$, that corresponds to the ground state of the anisotropic spin-($\frac{1}{2},\frac{3}{2}$) mixed Heisenberg chain.  For  $J_{\frac{1}{4}}(L,-1)<J<
J_{\frac{1}{6}}(L,-1)$, with $J_{\frac{1}{6}}(L,-1)
\approx 0.75$, the ground state will be given by the gauge sector  with $n=\frac{L}{4}$ 
values $g_i=\frac{15}{4}$,  and since among these gauge choices the  lowest energy is given by the most homogeneous 
distribution of $n=L/6$ values of $g_i=\frac{15}{4}$, 
the ground-state energy is the same as that of the mixed spin-($\frac{1}{2},\frac{1}{2},\frac{1}{2},\frac{3}{2}$) Heisenberg Hamiltonian where we have spin-$\frac{1}{2}$ operators in all lattice points except at the equally spaced points $i=4j$ ($j=1,2,\dots$).  For $J_{\frac{1}{6}}(L,-1)<J<
J_{\frac{1}{8}}(L,-1)$ the ground state energy should come from the gauge sector with the most homogeneous 
distribution 
with $n=\frac{L}{6}$ values of $g_i=\frac{15}{4}$, or equivalently  it is given by the mixed spin-($\frac{1}{2},\frac{1}{2},\frac{1}{2},\frac{1}{2},\frac{1}{2},\frac{3}{2}$) Heisenberg model.  Our numerical results induce us to expect\cite{conjec} that in fact for sufficiently large $L$, these crossings should continue, and we have in general for $J_{\frac{1}{2l}}(L,-1)<J<J_{\frac{1}{2(l+1)}}(L,-1)$
 ($l=1,2,3,\dots$) the ground-state energy in the sector with  $n=\frac{L}{2l}$ values of $g_i=\frac{15}{4}$, or equivalently it is given by the ground-state energy of the mixed spin-($\frac{1}{2}^{(2l-1)},\frac{3}{2}$) Heisenberg chain.  Fig. \ref{fig_diag} also shows that $(J_{\frac{1}{2(l+1)}}(L,-1)-
 J_{\frac{1}{2l}}(L,-1))\rightarrow 0$ as $l$ grows and for $J>J_0(L,-1)$ the ground state energy is given by the gauge $g_1=g_2=\dots=g_{\frac{L}{2}}=\frac{3}{4}$, that is equivalent to the ground-state energy of the exact solvable \cite{yangyang} standard antiferromagnetic spin-$\frac{1}{2}$ Heisenberg chain.  For other values of the anisotropy, as long $\Delta<1$, we have  the same qualitative behavior as $\Delta=-1$, with crossing values $J_{\frac{1}{2l}}(L,\Delta)$, however the crossing regions $\delta J^c(L,\Delta)=J_0(L,\Delta)-J_{1/2}(L,\Delta)$ decreases as we departure from $\Delta=-1$. Particularly for $\Delta\rightarrow -\infty$ or  $\Delta>1$  all the crossings 
happen at the same point $J_c(L,\Delta)$ ( $\delta J^c(L,\Delta) = 0$), and in those cases the ground-state energy is the same as that of the anisotropic mixed spin-($\frac{1}{2},\frac{3}{2}$) Heisenberg chain or the standard anisotropic spin-$\frac{1}{2}$ Heisenberg chain if $J<J_c(L,\Delta)$ or $J>J_c(L,\Delta)$,  respectively.

\section{The phase diagram of the tetrahedral spin chain}

In order to obtain the phase diagram of our Hamiltonian \eqref{hamiltonian} we need to estimate the bulk limit values $J_{\eta}(\infty,\Delta)={\lim}_{{L\rightarrow\infty}}J_{\eta}(L,\Delta)$ ($\eta=\frac{1}{2},\frac{1}{4},\frac{1}{6},
\dots$), where the ground-state change sectors and phase transitions are expected.  From the discussions of the last section and relation \eqref{E_finito} the crossing points for a finite lattice are given by
\begin{align}\label{J_eta}
J_{\eta}(L,\Delta)=\frac{2}{3}\frac{E_L^0(\eta L,\Delta,0)-E_L^0(\eta' L,\Delta,0)}{L(\eta'-\eta)},
\end{align}
with
\begin{align}
 \eta'=\frac{\eta}{1+2\eta},
\end{align}
where $\eta'=\frac{1}{4},\frac{1}{6},\frac{1}{8}, \dots,0$. 
Consequently the phase transition lines $J_{\eta}(\infty,\Delta)$ can be calculated 
from the finite size corrections of the energy levels $E_L^0(\eta L,\Delta,0)$ of the mixed  
spin-($\frac{1}{2},\frac{3}{2}$) Heisenberg model with $\eta L$ spins $\frac{3}{2}$ 
and $(1-\eta)L$ spins $\frac{1}{2}$ homogeneously distributed.

Based on general grounds the ground-state energy of these mixed Heisenberg models with 
a finite number $n=\eta L$ of spins $\frac{3}{2}$ should behave asymptotically ($L\rightarrow\infty$) as
\begin{align}
E_L^0(n,\Delta,0)=e_{\infty}^{\rm XXZ}(\Delta)L+F_L(n,\Delta),
\end{align}
where 
\begin{align}\label{E_surf}
F_L(n,\Delta)=F_{\infty}(n,\Delta)+ o(1/L),
\end{align}
and  $e_{\infty}^{\rm XXZ}(\Delta)$ is the ground-state energy per site of the anisotropic  spin-$\frac{1}{2}$ XXZ chain, with anisotropy $\Delta$, and $F_{\infty}(n,\Delta)$ plays the role of a surface energy (negative) due to the additional spin-$\frac{3}{2}$ particles.  In the region where the equivalent mixed Heisenberg chain is critical ($-1\leqslant\Delta\leqslant 1$), due to the conformal invariance of the infinite system, the neglected term in \eqref{E_surf} is ${\cal O}(1/L)$\cite{cardy_np,blote_cardy}, while in the non-critical regions it should behave 
as $e^{-\alpha L}$, where  $\alpha > 0 $ is proportional to the mass gap of the model. 
Moreover since the density of spin $\frac{3}{2}$  is bounded, i. e., $0\leqslant\eta=\frac{n}{L}\leqslant\frac{1}{2}$, 
we should expect in the limit where  $L\rightarrow\infty$, $n\rightarrow\infty$ but with 
$\eta=\frac{n}{L}$ kept fixed that the surface energy per site is given by 
\begin{align}
\frac{F_L(\eta L,\Delta)}{L} = f_{\infty}(\eta,\Delta)=\lim_{L\rightarrow\infty}\frac{F_L(\eta L,\Delta)}{L}.
\end{align}
where $f_{\infty}(\eta,\Delta)$ depends only on the value of $\eta$ and $\Delta$.

\begin{figure}[ht]
\begin{center}
 \psfrag{E_L(n)/L}[b][]{$f_{\infty}(\eta,-1)$}
 \psfrag{dE_L(n)/L}[b][]{$f'_{\infty}(\eta,-1)$}
 \psfrag{D=-1}[][]{$\Delta=-1$}
\psfrag{n/L}{$\eta$}
\psfrag{d2f_s}{$a_2(\Delta)$}
\includegraphics[width=5.5cm,height=7.7cm,angle=-90]{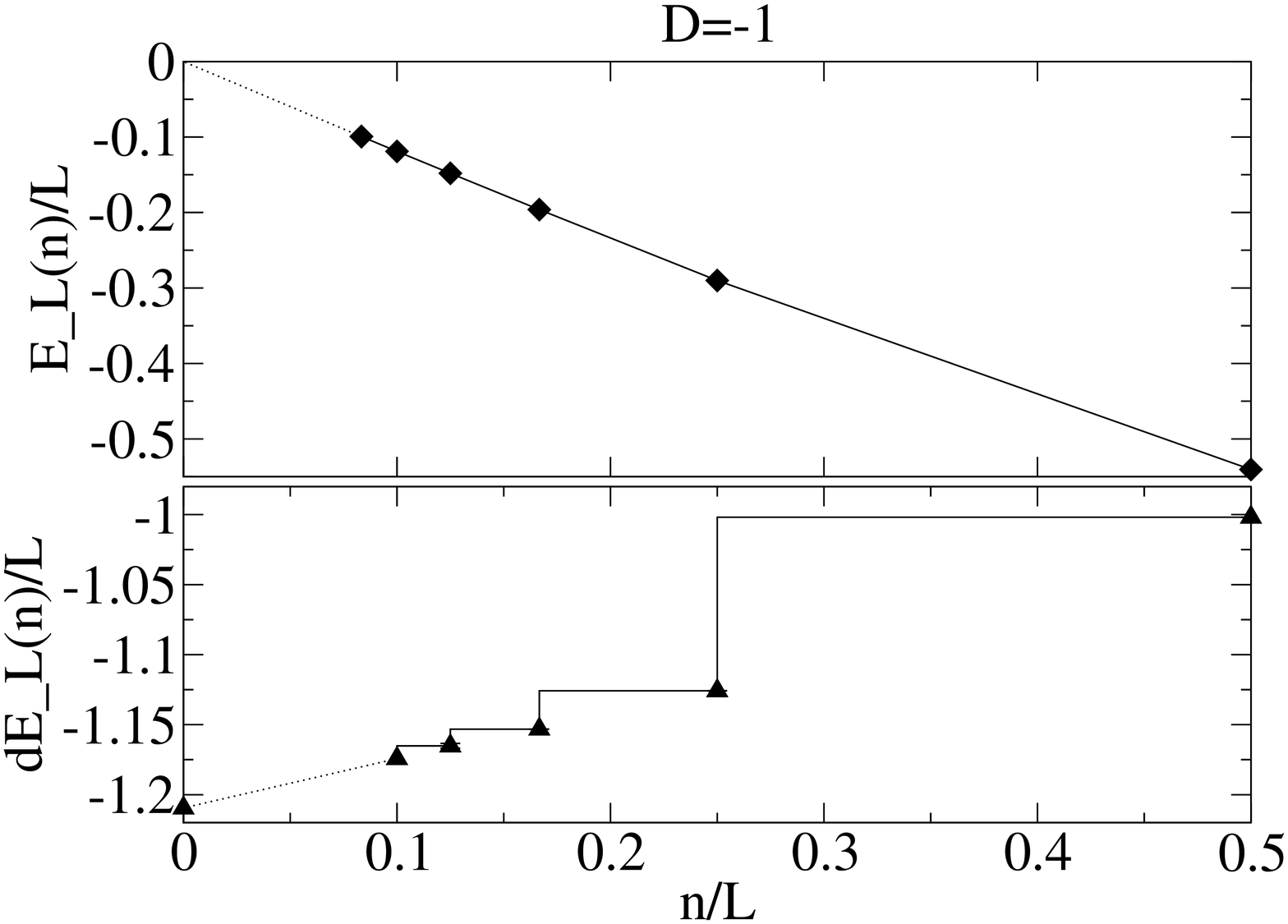}
\caption[fig2]{Surface energy per site  in the bulk limit $f_{\infty}(\eta,-1)$  
as a function of the density $\eta$ of spins $\frac{3}{2}$ (top) and its derivative 
$f'_{\infty}(\eta,-1)$ with respect to the density (bottom).}
\label{f_n}
\end{center}
\end{figure}

\begin{figure}[ht]
\begin{center}
 \psfrag{En-En-1}[b][][0.85]{$E^0_{L}(n,\Delta,0)-E_{L}^0(n-1,\Delta,0)$}
 \psfrag{D=-1}[][]{$\Delta=-1$}
\psfrag{n/L}{$n/L$}
\includegraphics[width=5.5cm,height=7.7cm,angle=-90]{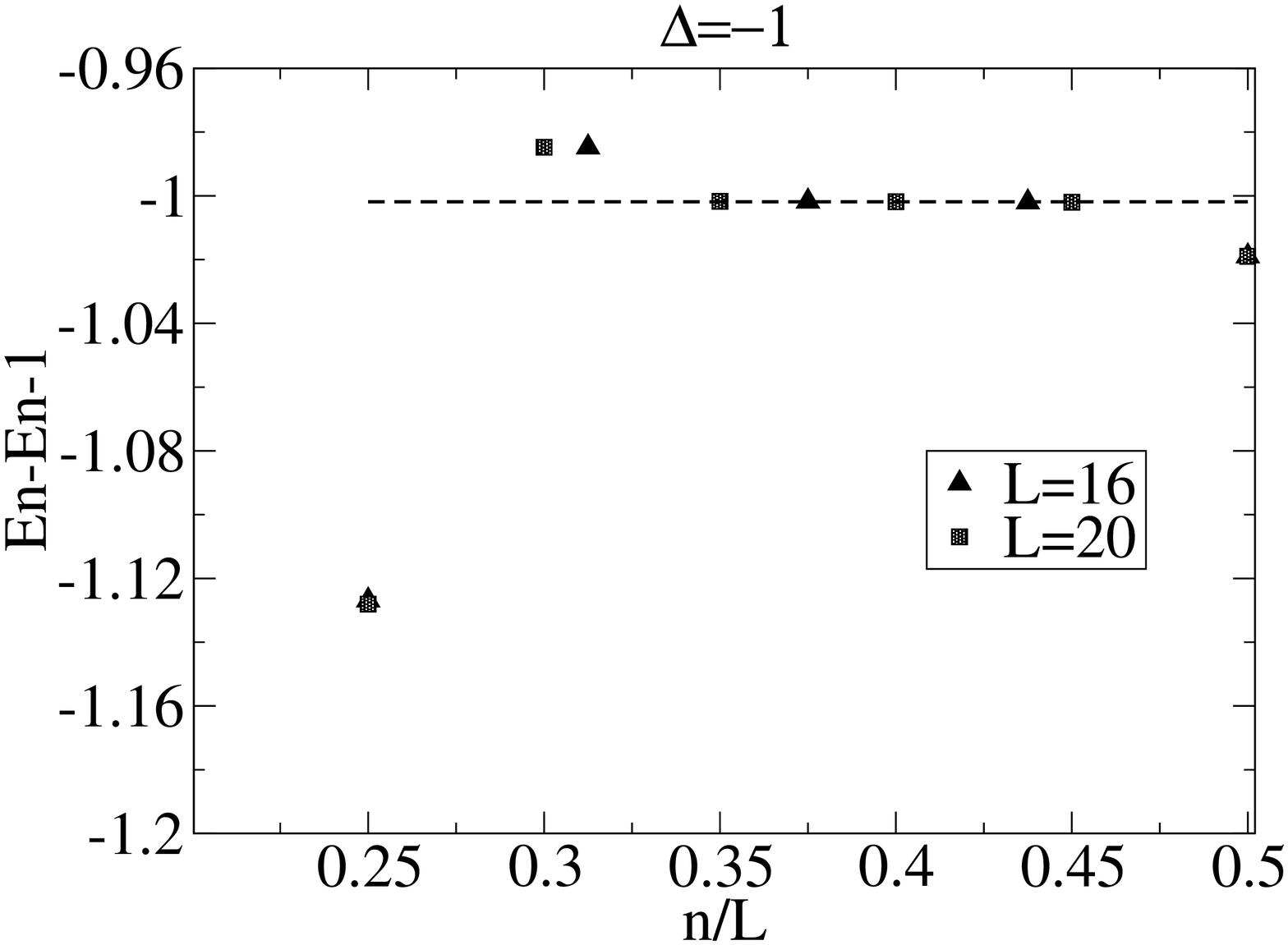}
\caption[fig2]{Difference among the two lowest energies $E_L^0(\eta L,-1,0)$ and $E_L^0(\eta L -1,-1,0)$ for 
the lattice sizes $L=16$ (triangles) and $L=20$ (squares) in the gauge 
sector where the  density $\eta = n/L$ of spins $\frac{3}{2}$ changes from 0.20 to 0.5. The dashed line  represents the bulk limit 
in the intermediate region $\frac{1}{4} < \eta< \frac{1}{2}$.}
\label{f_n_cons}
\end{center}
\end{figure}

The  finite-size extrapolations
$f_{\infty}(\eta,\Delta)$ for the ferrimagnetic case ($\Delta = -1$) for some 
densities are shown (diamonds)  in Fig. \ref{f_n}.  Although the curve formed 
by the union of these points is almost a straight line there exists a small 
decreasing upward concavity as $\eta$ decreases. This fact will be crucial 
to explain the existence of an infinite number of phase transitions for 
the model around $\Delta \sim -1$. 
In order to show these  concavities  let us consider the densities in the
region 
$\frac{1}{4}<\eta<\frac{1}{2}$. The largest lattice we can calculate with 
the most homogeneous distribution, for these densities is  
$L=16$ (triangles) and $L=20$ (circles). In Fig. \ref{f_n_cons} by 
considering these lattice sizes we show the finite-size version of these 
derivatives, namely, the mass gap obtained by the increment of a single 
impurity in the system, i. e., $E^0_L(\eta L,-1,0) - E^0_L(\eta L-1,-1,0)$. 
 This figure indicates that except for the densities close to $\eta=\frac{1}{4}$
 or $\eta = \frac{1}{2}$ this gap has the constant value $G_1 = -1.0019$.  The
 same behavior should also happens for the 
 densities in other intervals $\frac{1}{2(l+1)} < \eta < \frac{1}{2l}$ 
 ($l=2,3,\ldots$), where the gap has an almost constant value $G_l$ that 
 decreases as $l$ increases. The small deviations shown in Fig. \ref{f_n_cons} 
 for $G_1$ around $\eta=0.25$ and $\eta=0.5$ are due to the finite-size 
 effects.


In the lower part of Fig. \ref{f_n} we plot the numerical values obtained for
$G_l$ ($l=1,2,\ldots$). A similar analysis for other values of $\Delta$ show us
that except for $\Delta >1$ (ferromagnetic region) a similar behavior happens. 
These results imply that in the bulk limit we should expect that the 
surface energy per site is given by
\begin{align}\label{12p}
f_{\infty} (\eta,\Delta) = \eta G_l(\Delta), \quad l = \mbox{Int}[1/2\eta].
\end{align}

\begin{figure}[ht]
\begin{center}
 \psfrag{D=-1}[][]{$\Delta=-1$}
 \psfrag{E1-E0}[b][][0.85]{$E_L^0(1,\Delta,0)-E_L^0(0,\Delta,0)$}
\psfrag{1/L}[][]{1/L}
\includegraphics[width=5.5cm,height=7.8cm,angle=-90]{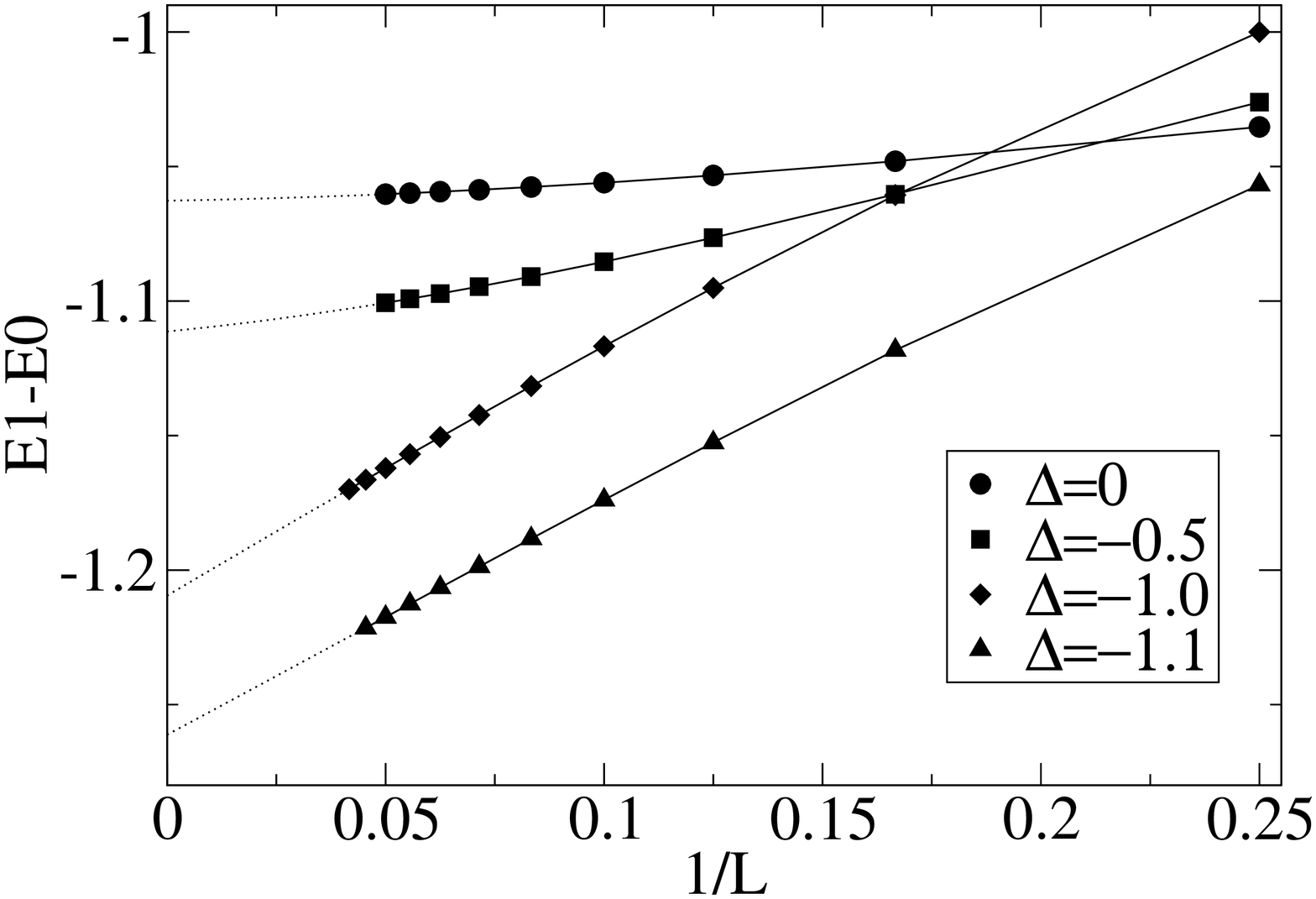}
\caption[fig2]{Extrapolation of the difference among the 
ground-state energy of the homogeneous anisotropic 
spin-$\frac{1}{2}$ Heisenberg chain and the spin-$\frac{1}{2}$ 
Heisenberg chain with a single  impurity of spin $\frac{3}{2}$.}
\label{E1-E0}
\end{center}
\end{figure}

In the limit of lower density ($\eta \sim 0$) and higher density ($\eta \sim
\frac{1}{2}$)
we can obtain, as a function of $\Delta$,  accurate results  for
$G_{\infty}(\Delta)$ and $G_1(\Delta)$, respectively.  The results for
$\eta\sim 0$
($\eta\sim \frac{1}{2}$) are calculated from lattice sizes up to $L=24$ 
($L=20$), where we
have ($L-1$) spins $\frac{1}{2}$ (spins $\frac{3}{2}$) with a single impurity
of spin $\frac{3}{2}$ (spin $\frac{1}{2}$). In Fig. \ref{E1-E0} the
finite-size sequences are shown for the surface energy in the lower density
limit $\eta=0$, and for some values of $\Delta$. Using a cubic least-square 
fitting we obtained the extrapolated values of $G_{\infty}(\Delta)$ shown in
table \ref{Tab:1}, where the errors are estimated from the stability of the 
fitting when the largest lattices is neglected. A similar extrapolation for 
$\eta \sim\frac{1}{2}$ shows that $G_1(\Delta) \sim -1.00187$ for all values of 
$\Delta$ in the critical region. In the massive regions $|\Delta| >1$ we have
two distinct behaviors, for $\Delta \ll -1$,  $G_1(\Delta) = +\Delta$ while for
$\Delta>1$ (ferromagnetic region) $G_l(\Delta) = -\Delta$, for all values of
$l$. This means that in the massive ferromagnetic regime 
$f_L(\eta,\Delta) =
-\eta\Delta$.

The surface energy \eqref{12p} when inserted in \eqref{E_finito} give us the
bulk value ($n\rightarrow \infty, L\rightarrow \infty, 
\eta= n/L$ finite) for the energy per site of the lowest 
eigenenergy in the gauge sector with $n = \eta L$ values of $g_i =
\frac{15}{4}$
\begin{widetext}
\begin{align} \label{12pp}
e_{\infty} (\eta,\Delta,J) = \mbox{lim}_{L\rightarrow \infty}
\frac{E_L^0(\eta L,\Delta,J)}{L} = 
e_{\infty}^{XXZ} (\Delta) + G_l(\Delta)\eta - \frac{3}{4} 
(\frac{1}{2}-2\eta)J.
\end{align}
\end{widetext}

This last relation implies that the  critical couplings $J_{\eta}(\infty,\Delta)$,  
given by  \eqref{J_eta},  where the ground state changes its location on the 
gauge sectors  are just 
given by 
\begin{align} \label{12ppp}
J_{\eta}(\infty,\Delta) = -\frac{2}{3}G_{\frac{1}{2\eta}} (\Delta), \quad 
\eta = \frac{1}{2},\frac{1}{4},\frac{1}{6},\dots
\end{align}

\begin{table}[ht]
\centering
\begin{tabular}{|l|l|c|l} \hline\hline
 {$\Delta$ } & {\; $G_ l(\Delta)$\; } & { \;$\delta G_l(\Delta)$ \;}\\  \hline\hline
$1$  & -1.0000 & 0   \\ 
$\tfrac{\sqrt 3}{2}$ & -1.0067 & -0.0001\\
$\tfrac{\sqrt 2}{2}$ & -1.0162 & -0.0002\\
$0.5$                & -1.0289 & -0.0002\\ 
0                    & -1.0627 &  -0.0002 \\ 
-$0.5$                & -1.1113 &  -0.0004 \\
-$\tfrac{\sqrt 2}{2}$ & -1.1425 &  -0.0004\\ 
-$\tfrac{\sqrt 3}{2}$ & -1.1746 &  -0.0001\\
-1               &     -1.2095 & 0.0008 \\
-1.1             &     -1.2612 & 0.0005 \\
-1.2             &     -1.3250 & 0.0004 \\
-1.3             &     -1.3996 & 0.0005 \\
-1.5             &     -1.5734 & 0.0009 \\
-2.0             &     -2.0539 &  -0.0054\\\hline 
\end{tabular}
\caption{\label{Tab:1} The energy gaps are tabulated at very low density 
($1\ll l < \infty$) of concentration in the critical and massive antiferromagnetic regions.}
\end{table}

\begin{widetext}

\begin{figure}[ht]
\begin{center}
 \psfrag{J}[]{$J$}
 \psfrag{D}[]{$\Delta$}
\psfrag{ferrm1}{$\jn\jn\jn\jn\jn\jn$}
\psfrag{ferrm2}{$\Big\jn\!\jn\!\Big\jn\!\jn\!\Big\jn\!\jn$}
\psfrag{antifm1}{$\jn\ur\jn\ur\jn\ur$}
\psfrag{antifm2}{$\Big\jn\!\ur\!\Big\jn\!\ur\!\Big\jn\!\ur$}
\psfrag{disorder_phase1}{\small{critical disorder ($\frac{1}{2},\frac{1}{2}$)}}
\psfrag{disorder_phase2}{\small{critical disorder ($\frac{1}{2},\frac{3}{2}$)}}
\psfrag{trans1}{\small{$J_{\delta\eta\rightarrow 0}$}}
\psfrag{trans2}{\small{$J_{\frac{1}{2}\rightarrow\frac{1}{4}}$}}
\psfrag{trans3}{\small{$J_{\frac{1}{4}\rightarrow \frac{1}{6}}$}}
\psfrag{A11}{$A_{\frac{1}{2},\frac{1}{2}}$}
\psfrag{D11}{$D_{\frac{1}{2},\frac{1}{2}}$}
\psfrag{F11}{$F_{\frac{1}{2},\frac{1}{2}}$}
\psfrag{A13}{$A_{\frac{1}{2},\frac{3}{2}}$}
\psfrag{D13}{$D_{\frac{1}{2},\frac{3}{2}}$}
\psfrag{F13}{$F_{\frac{1}{2},\frac{3}{2}}$}
\includegraphics[width=8cm,height=14cm,angle=-90]{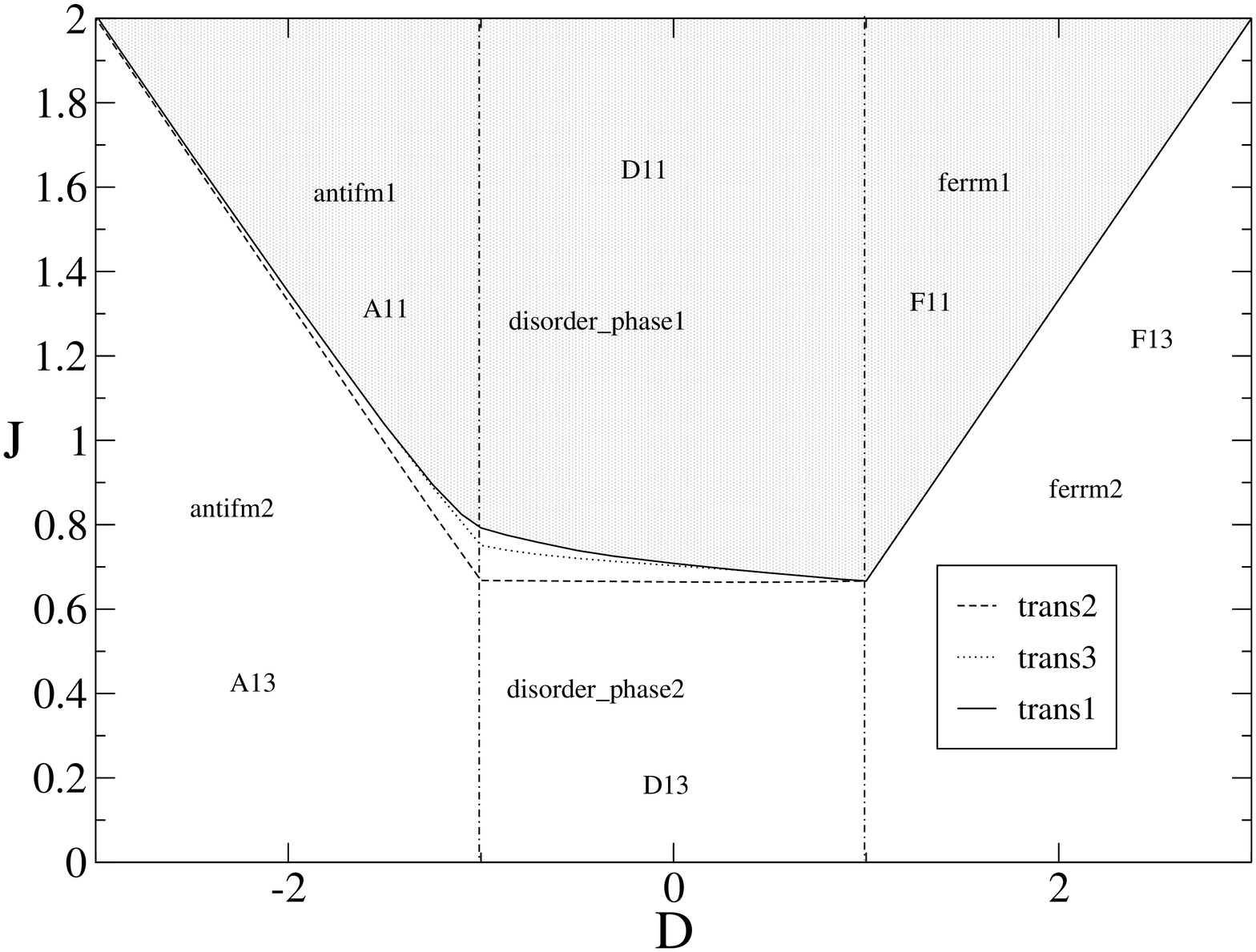}
\caption[fig_DJ]{Phase diagram at zero temperature of the tetrahedral 
spin chain with Hamiltonian \eqref{hamiltonian}.}
\label{fig_DJ}
\end{center}
\end{figure}

\end{widetext}

In Fig. \ref{fig_DJ} we plot for the densities $\eta=0,\frac{1}{4},\frac{1}{2}$ these lowest energies as a function of $J$ for $\Delta=-1$, $\Delta=0$  and $\Delta=1$. We see from this figure that while for $\Delta=-1$ and $\Delta=-0.5$ there exist  crossings from the curves with $\eta=0$ and $\eta=\frac{1}{6}$, these do not happen for $\Delta>1$, in agreement with our discussions with finite size lattice (see Fig. \ref{fig_diag}). It is important to stress here that those crossings are governed by the intercept with the vertical axis
$e_{\infty}(\eta,\Delta)$, whose $\Delta$-dependence in the bulk limit are given by the ground-state energy of the XXZ chain and the surface energies $f_{\infty}(\eta,\Delta)$.  The exact results of $e_{\infty}^{\rm XXZ}(\Delta)$ \cite{yangyang} and our numerical estimates for $G_l(\Delta)$ show us the following asymptotic behavior. For $\Delta>1$ there is a single crossing from the density $\eta=\frac{1}{2}$  to $\eta=0$ at $\Delta=\frac{3J}{2}$, where the ground state changes from that of the mixed spin-($\frac{1}{2},\frac{3}{2}$) Heisenberg chain to the standard  spin-$\frac{1}{2}$ XXZ chain.  For $\Delta<1$ there exist level crossings for all the gauge choices with $\eta=\frac{1}{2l}$, $l=1,2,\dots$. The range of values of $J$ where these crossings happen is relatively small except around $\Delta\approx -1$.

Collecting our numerical results we obtain the quite rich phase diagram shown in Fig. \ref{fig_DJ} for our tetrahedral spin chain.  The phase diagram shows eight distinct phases that we are going to describe.  The ground state energy in the phases $A_{\frac{1}{2},\frac{1}{2}}$, $D_{\frac{1}{2},\frac{1}{2}}$,  $F_{\frac{1}{2},\frac{1}{2}}$ (shadow region in Fig. \ref{fig_DJ}) is given, apart from degeneracies, by the ground-state energy of the standard spin-$\frac{1}{2}$  XXZ chain, while the phases $A_{\frac{1}{2},\frac{3}{2}}$, $D_{\frac{1}{2},\frac{3}{2}}$,  $F_{\frac{1}{2},\frac{3}{2}}$ are given by the ground-state of the anisotropic mixed spin-(${\frac{1}{2},\frac{3}{2}}$) Heisenberg chain. The intermediate phases $A_{\frac{1}{2},\eta}$ and $D_{\frac{1}{2},\eta}$, actually are regions with an infinite 
 sequence of  phase transitions where the ground-state energy is that of the anisotropic version of a mixed Heisenberg chain with spins $\frac{1}{2}$ and $\frac{3}{2}$, where the density of spin $\frac{3}{2}$ is $\eta=\frac{1}{2},\frac{1}{4},\frac{1}{6},\dots,0$.  Despite of the large number of closed excited states on each of these phases the ground state is given by a distinct wave function with peculiar magnetic properties. The phase diagram of Fig. \ref{fig_DJ} can be divided in three distinct  regions with respect to the variations of  $\Delta$.

\begin{itemize}
\item[{\it i)}] $\Delta\geqslant 1$. For $J>\frac{2\Delta}{3}$ (phase $F_{\frac{1}{2},\frac{1}{2}}$)  the system behaves  as 
a ferromagnetic Heisenberg spin-$\frac{1}{2}$ chain with  $z$-magnetization $M^z=\pm \frac{L}{2}$ while for $J<\frac{2\Delta}{3}$ (Phase $F_{\frac{1}{2},\frac{3}{2}}$) it behaves effectively as a mixed spin-($\frac{1}{2},\frac{3}{2}$) Heisenberg chain in its two fully ordered state with total  $z$-magnetization $M^z=\pm L$. At $J=\frac{2}{3}\Delta$ there is a first order phase transition separating these two phases. At $\Delta=1$ the phases  $F_{\frac{1}{2},\frac{1}{2}}$ and $F_{\frac{1}{2},\frac{3}{2}}$ end up into a gapless phase transition line where all the lowest energies in the sectors with $z$-magnetization $-M^z<m^z<M^z$ degenerate, rendering a ferromagnetic ordered state with total magnetization $M^z$ and a massless spectra. When $\Delta\rightarrow\infty$ (Ising limit) the ground-state energy per site in 
the phase $F_{\frac{1}{2},\frac{1}{2}}$ is ${\mathcal E}_0=-\frac{\Delta}{4}-\frac{3J}{8}$ and in the phase $F_{\frac{1}{2},\frac{3}{2}}$ is ${\mathcal E}_0=-\frac{3\Delta}{4}+\frac{3J}{8}$
\item[{\it ii)}] $|\Delta|<1$.  For $J<\frac{2}{3}$ the system behaves as the disordered phase of the mixed spin-($\frac{1}{2},\frac{3}{2}$) Heisenberg chain.  For $J>J_0(\infty,\Delta)$ (shadow region) the model behaves as the standard and solvable XXZ chain in its massless regime.  For $\frac{2}{3}=
J_{\frac{1}{4}}<J<J_0(\infty, \Delta)$ there exists an infinite number of phase transition separating the regions $J_{\eta'}(\infty,\Delta)<J<J_{\eta}(
\infty,\Delta)$ ($\eta=\frac{1}{2l}$, $\eta '=\frac{1}{2(l+1)}$, $l=2,3,\dots$) where the model behaves effectively as a mixed spin-(($\frac{1}{2})^{2l-1},\frac{3}{2}$) Heisenberg chain with density $\eta=\frac{1}{2l}$  ($l=2,3,\dots$) of spins $\frac{3}{2}$, in its massless regime.  
These intermediate phases happen in a small range of 
values of $J$ in the phase diagram, 
except around $\Delta=-1$.  At $\Delta=-1$ 
all these phases end up into a phase transition line  
where the system try to establish an antiferromagnetic order. 
This antiferromagnetic order is achieved in the phase  
$D_{\frac{1}{2},\frac{1}{2}}$.  
However in all the phases $D_{\frac{1}{2},\eta}$ 
and  $D_{\frac{1}{2},\frac{3}{2}}$, due to the Lieb-Mattis theorem 
\cite{liebmattis}, the lowest eigenenergy in  several sectors with 
distinct magnetizations (not all of them)
are 
degenerated and we have phases with 
 increasing ferrimagnetic order.

\item[{\it iii)}] $\Delta<-1$.  For $J<\frac{2\Delta}{3}$ (phase $A_{\frac{1}{2},\frac{3}{2}}$) the model behaves as an effective spin ($\frac{1}{2},\frac{3}{2}$) Heisenberg model with  massive ferrimagnetic behavior, 
while for $J>J_0(\infty,\Delta)$ (phase $A_{\frac{1}{2},\frac{1}{2}}$) the model behaves as the antiferromagnetic spin-$\frac{1}{2}$ Heisenberg model.  For the intermediate values $\frac{2\Delta}{3}<J<J_{\eta}(\infty,\Delta)$ the model exhibits a sequence of ferrimagnetic ordered phase, behaving effectively as a mixed spin-($\frac{1}{2},\frac{3}{2}$) Heisenberg model with distinct densities of spins $\frac{3}{2}$.
\end{itemize}

It is interesting to stress here that in all the shadow region of Fig. \ref{fig_DJ} the ground state have a degeneracy of order $2^{\frac{L}{2}}$ in a $L$-sites chain, due to the possible gauge choices of our model \eqref{hamiltonian}. 
This means that similarly as happen with the ice at low temperatures
 \cite{pauling} the entropy 
 per site in all these regions is the finite number $s_o
= \frac{\kB}{2} \ln 2$, where $\kB$ is the Boltzmann constant.

\section{Thermodynamics}

In this section we discus the thermodynamic properties of the  tetrahedral spin chain \eqref{hamiltonian}. The gauge 
symmetry of \eqref{hamiltonian} enable us to write its partition function 
(see \eqref{d0})
\begin{widetext} 
\begin{align}\label{d0}
{\cal Z}_L(T,\Delta,J)=\big({\rm 2e}^{\frac{3J}{4\kB T}}\big)^{\frac{L}{2}}\sum_{n=0}^{\frac{L}{2}}
\Big({\frac{e^{-\frac{3J}{2\kB T}}}{2}} \Big)^n 
\sum_{1\leq i_1 <i_2 <\cdots<i_n\leq L} 
Z_L(i_1,\ldots,i_n; T,\Delta),
\end{align}
where $Z_L(i_1,\ldots,i_n;T,\Delta)$ is the partition function of the spin-$\frac{1}{2}$ anisotropic Heisenberg chain 
with $n$ spin-$\frac{3}{2}$ impurities located at $i_1,\ldots,i_n$, and $\kB$,
as before,  is the Boltzmann constant. 

As $T \rightarrow 0$ we can approximate 
\begin{align}\label{d1}
Z_L(i_1,\ldots,i_n;T,\Delta) \simeq g_L^0
(i_1,\ldots,i_n)e^{-\frac{E^0(i_1,\ldots,i_n;\Delta)}{\kB T}},
\end{align}
where $E^0(i_1,\ldots,i_n;\Delta)$ is the ground-state energy of the equivalent
spin-$\frac{1}{2}$ Heisenberg chain with impurities at $\{i_1,\ldots,i_n\}$,
whose degeneracy is given by $g_L^0(i_1,\ldots,i_n)$. Even with the
approximation \eqref{d1} the evaluation of $Z_L(T)$ is not simple for general
values of $J$ and $\Delta$. In the ferromagnetic regime $\Delta  \geq 1$ however
some 
additional simplifications occur. As discussed in \S 2 the degeneracy
$g^0_L(i_1,\ldots,i_n) =2$ and the ground-state energy $E_L^0(i_1,\ldots,i_n;\Delta)$
does not depend on the specific locations of the impurities, only on their
number. Moreover as $L\rightarrow \infty$, taking into account that for 
$\Delta >1$, $G_l(\Delta) = -\Delta$ in \eqref{12p}-\eqref{12pp} we have 
\begin{align} \label{d2}
E_L^0(i_1,\ldots,i_n;\Delta) = E^0(n,\Delta) = 
Le_{\infty}^{XXZ}(\Delta) - n\Delta .
\end{align}
Inserting \eqref{d2} and \eqref{d1} in \eqref{d0} we obtain
\begin{align} \label{d3}
Z_L(T,\Delta,J) = 2{\rm e}^{-\frac{e_{\infty}^{XXZ}(\Delta)L}{\kB T}}\Big[
2{\rm e}^{\frac{3J}{4\kB T}} + {\rm e}^{(\Delta+\frac{3}{4}J)/\kB T}\Big]^{\frac{L}{2}},
\end{align}
for $T\rightarrow 0$, $L\rightarrow \infty$ and $\Delta >1$.
This expression gives the entropy per particle
\begin{align} \label{d4}
s(T,\Delta,J) = \frac{\kB}{2} \ln (2+{\rm e}^{-\frac{3J-2\Delta}{\kB T}} ) + 
\frac{1}{2T} \frac{{\rm e}^{-\frac{3J-2\Delta}{\kB T}}(3J-2\Delta)}
{2 +{\rm e}^{-\frac{3J-2\Delta}{\kB T}}}.
\end{align}
\end{widetext}

\begin{figure}[ht]
\begin{center}
\psfrag{S}[b][][0.95]{$s(T,\Delta,J)/\kB$}
\psfrag{J}[][]{\small$J$}
\psfrag{T=0.001}[][]{\small$T=0.001/\kB$}
\includegraphics[width=5.5cm,height=7.7cm,angle=-90]{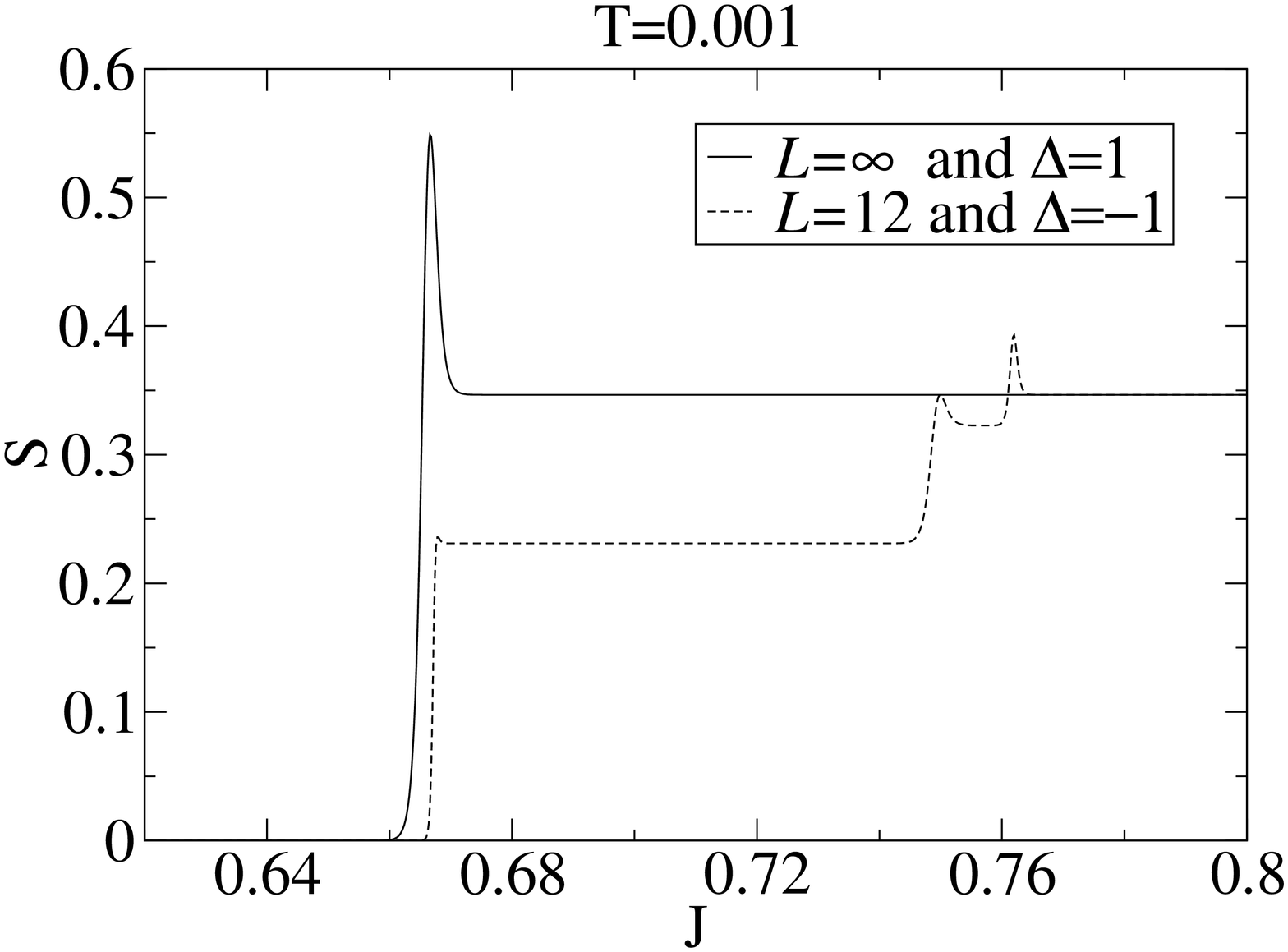}
\caption[fig2]{Entropy at low temperatures in the ferromagnetic regime ($\Delta=1,L\rightarrow\infty$) and in the ferrimagnetic regime ($\Delta=-1,L=12$).}
\label{Low_T_S}
\end{center}
\end{figure}

Exactly at $T=0$ we have $s(T,\Delta,J) =0$ for $J<J_c^0=\frac{2}{3}\Delta$ and 
$s(T,\Delta,J) = \frac{\kB }{2}\ln 2$ for $J>J_0(\infty,\Delta)$.
The expression \eqref{d4}  should be valid   
		for $\kB T \lesssim 0.1$, since the gap in the ferromagnetic regime, 
		that controls the approximation \eqref{d1}, is  of order
of 0.1 for the XXZ chain with an arbitrary number of impurities. In figure  
\ref{Low_T_S} we show the entropy per site given by \eqref{d4} at $T=0.001\kB $ 
and $\Delta = 1$. We see from this figure that the net effect of the temperature
is to produce a peak at the zero temperature quantum critical point
$J_0(\infty,\Delta)$.

\begin{figure}[ht]
\begin{center}
\psfrag{C}[b][][0.9]{$C_L(T,\Delta,J)/\kB$}
\psfrag{J}[t][]{\small$J$}
\psfrag{D=1 and T=0.01}[][]{\small$\Delta=1$ and $T=0.01/\kB$}
\psfrag{D=-1 and T=0.005}[][]{\small$\Delta=-1$ and $T=0.005/\kB$}
\includegraphics[width=5.5cm,height=7.7cm,angle=-90]{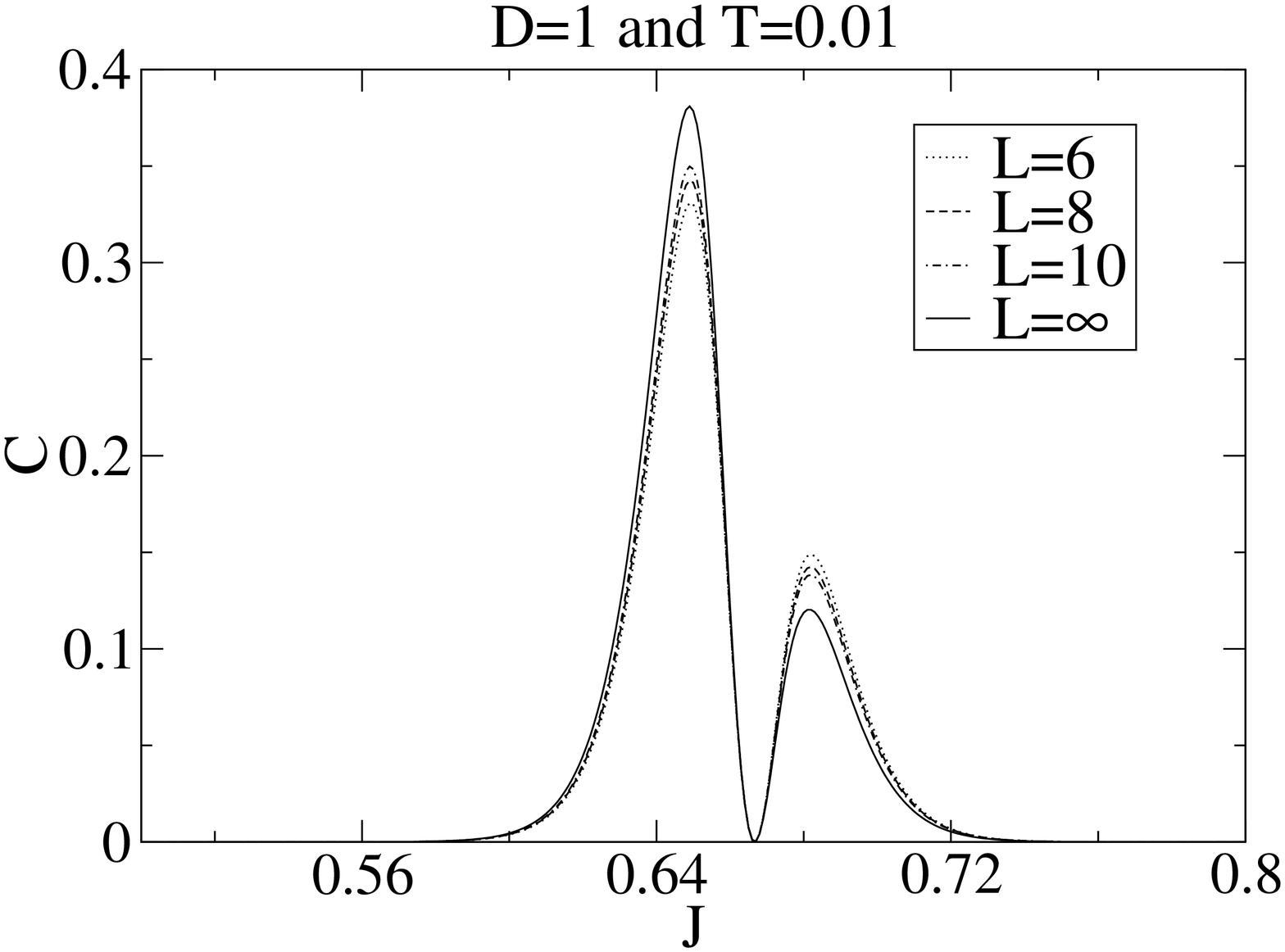}
\includegraphics[width=5.5cm,height=7.7cm,angle=-90]{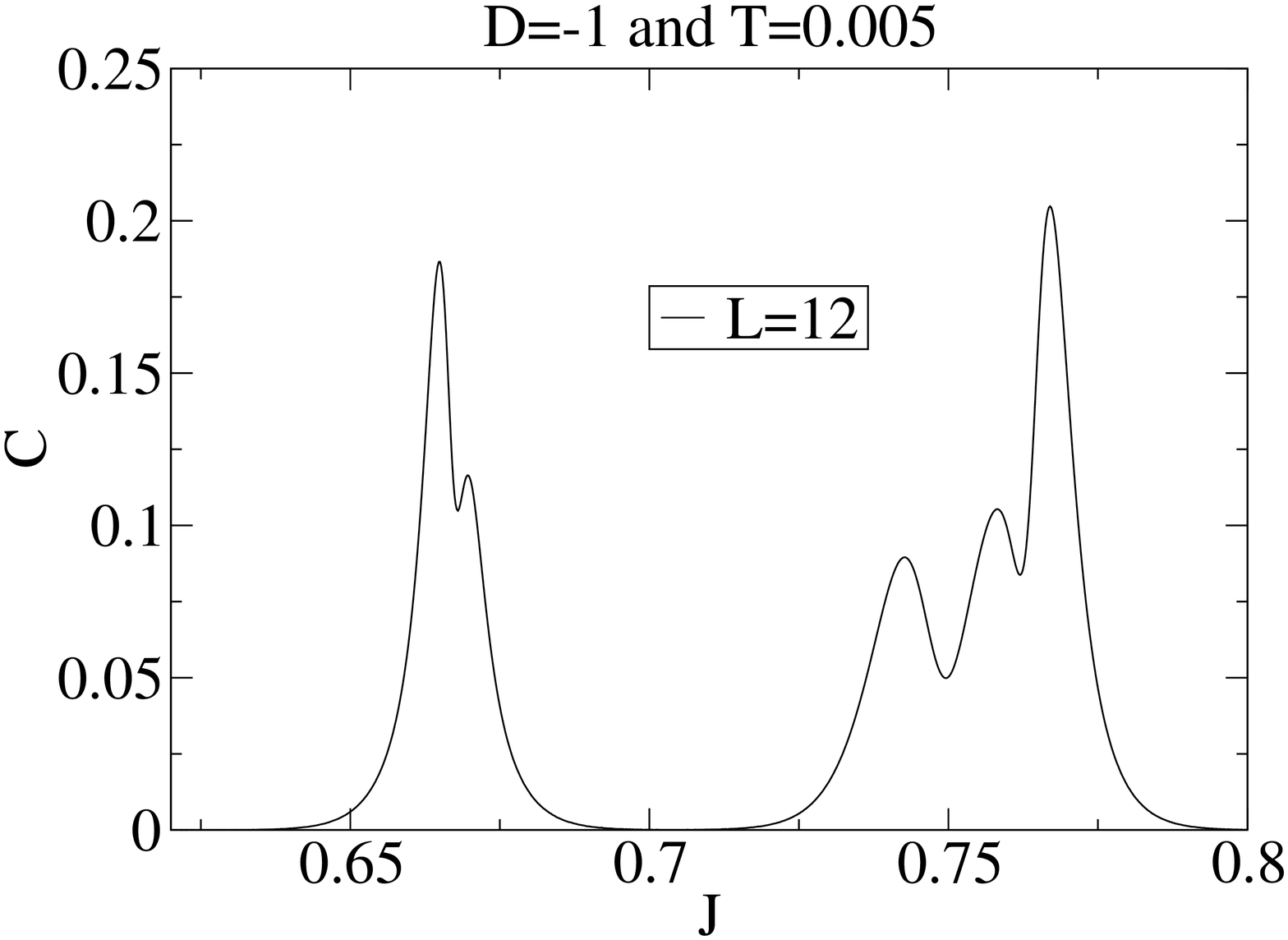}
\caption[fig2]{The specific heat at very low temperatures for the finite rings $L=6,8,10$  in the ferrimagnetic ($\Delta=-1$) region and in the ferromagnetic region ($\Delta=1$) with $L=10$ sites.}
\label{Low_T_Csp}
\end{center}
\end{figure}

The entropy for the other anisotropies $\Delta <1$ can only be estimated 
numerically from the exact numerical calculations on small lattices. We show in 
figure \ref{Low_T_S}, in order to compare with the ferromagnetic case 
$\Delta =1$, given by \eqref{d4}, the numerical results for the lattice $L=12$
of the ferrimagnetic model at $\Delta =-1$ and $T = 0.001/\kB $. In this case 
the entropy shows small peaks around the zero temperature phase 
transition couplings $J_{\frac{1}{2}}(\infty,\Delta), J_{\frac{1}{4}}(\infty,\Delta),\ldots$ . Strictly at 
$T=0$ and for $-1 \leq \Delta <1$ these peaks disappear and the entropy 
per site is given by 
the set of plateaux 
\begin{widetext}
\begin{align} \label{d44}
s(0,\Delta,J) =\frac{\kB}{2}(1 + \frac{1}{l}) \ln 2, \quad
J_{\frac{1}{2l}}(\infty,\Delta)<J<J_{\frac{1}{2(l+1)}}(\infty,\Delta). 
\end{align}

As we can see in the phase diagram of Fig. \ref{fig_DJ}  for $\Delta<1$ the
sizes of the plateaux $(J_{\frac{1}{2(l+1)}}(\infty,\Delta) - 
J_{\frac{1}{2l}}(\infty,\Delta)) \rightarrow 0$,
for any value of $\Delta< 1$, while for $\Delta \geq 1$ 
and $\Delta \ll -1$ there exists a single
plateaux starting at $J= J_0(\infty,\Delta) = \frac{2}{3}\Delta$.

The evaluation of the partition function \eqref{d1} for arbitrary temperatures
can be done by collecting all the eigenvalues of the several XXZ quantum chains 
with arbitrary distribution of spin-$\frac{3}{2}$ impurities at the even sites 
of the lattice. We can perform these calculations up to the lattice size $L=10$. In
order to extend these calculations for $L=12$ we would need to diagonalize 
matrices of order $32424\times 32424$, which is beyond the computational 
possibilities nowadays.

\subsection{The specific heat}

Let us consider initially the specific heat  at very low temperature. In the
massive ferromagnetic case we have from \eqref{d4} 
\begin{align} \label{d5}
C_L(T,\Delta,J) = \kB\Big(\frac{3J-2\Delta}{\kB T}\Big)^2\Big[ 2{\rm e}^{-\frac{3J-2\Delta}{4\kB T}} + 
{\rm e}^{\frac{3J-2\Delta}{4\kB T}}\Big]^{-2},
\end{align}
\end{widetext}
for $L\rightarrow \infty$, $\kB T \lesssim 0.1$ and $\Delta >1$. In order to
have an idea of the finite size-effects in the low temperature regime, where 
they are more important, we plot in Fig. \ref{Low_T_Csp} the bulk 
limit ($L\rightarrow \infty$) of the specific heat at $\Delta =1$ and 
$\kB T = 0.01$, given by \eqref{d5} (solid line) together with its finite size 
versions for lattice sizes $L=6,8$ and $10$ 
(dotted, dashed and dotted-dashed lines respectively), obtained by a direct numerical
diagonalization of the quantum chains.
These curves show that the main features of the specific heat for the
ferromagnetic chains at $T = 0.01$ are already shown for relatively 
small lattices. From \eqref{d5} for $L\rightarrow \infty$, $\Delta >1$ and 
$\kB T \lesssim 0.1$ two peaks happen at 
$J = J^{(1)} = \frac{2}{3}\Delta - 1.769772\kB T$ and $J=J^{(2)} = 
\frac{2}{3}\Delta  + 1.485203 \kB T$ with intensities 
$C_{\infty}(T,\Delta,J^{(1)}) = 0.388090112\kB $ and 
$C_{\infty}(T,\Delta,J^{(2)}) = 0.12038896 \kB $. It is interesting to observe 
that for these low temperatures the intensity of these peaks does not depend
on the temperature and they happen nearly at  the zero temperature 
phase transition point $J_0(\infty,\Delta) = \frac{2}{3}\Delta$. 
Also we can see from \eqref{d5} that the peaks at $J^{(1)}$ and $J^{(2)}$ are
due to the temperatures fluctuations of the ferromagnetic XXZ chain levels and 
the ferromagnetic mixed spin chain, respectively. Strictly at $T=0$ these two
peaks add up at the phase transition point 
$J= J_0(\infty,\Delta) = \frac{2}{3}\Delta$.

In the case of the non-ferromagnetic regime ($\Delta <1$), even in the 
low temperature region, an analytical calculation of the partition function, 
or the specific heat, is quite complicated. In order to compare with the 
ferromagnetic case we plot in Fig. \ref{Low_T_Csp} the specific heat 
of the ferrimagnetic chain $\Delta = -1$ at $\kB T = 0.0005$ as a function 
of $\Delta$, for the lattice size $L=12$. This calculation was performed  
directly from the  partition function \eqref{d0} obtained by considering only the lowest eigenenergy of the several eigensectors.   This lattice size give us a good finite-size realization 
for the densities $\eta = \frac{1}{2}, \frac{1}{4}$ and $\frac{1}{6}$,  and 
comparing with similar results for $L=16$ (not plotted in Fig. \ref{Low_T_Csp})
we verify that the main features of the $L\rightarrow \infty$ system is 
already kept with this lattice size. We see that as in the ferromagnetic case 
the peaks happen around the zero temperature phase transition points (see 
the phase diagram of Fig. \ref{fig_DJ}).
 
\begin{figure}[ht]
\begin{center}
\psfrag{C}[b][][0.9]{\small$C_L(T,\Delta,J)/\kB$}
\psfrag{T}[t][]{\small$\kB T$}
\psfrag{D=-1}[][]{$\Delta=-1$}
\psfrag{D=-10}[][]{$\Delta=-10$}
\includegraphics[width=5.5cm,height=7.7cm,angle=-90]{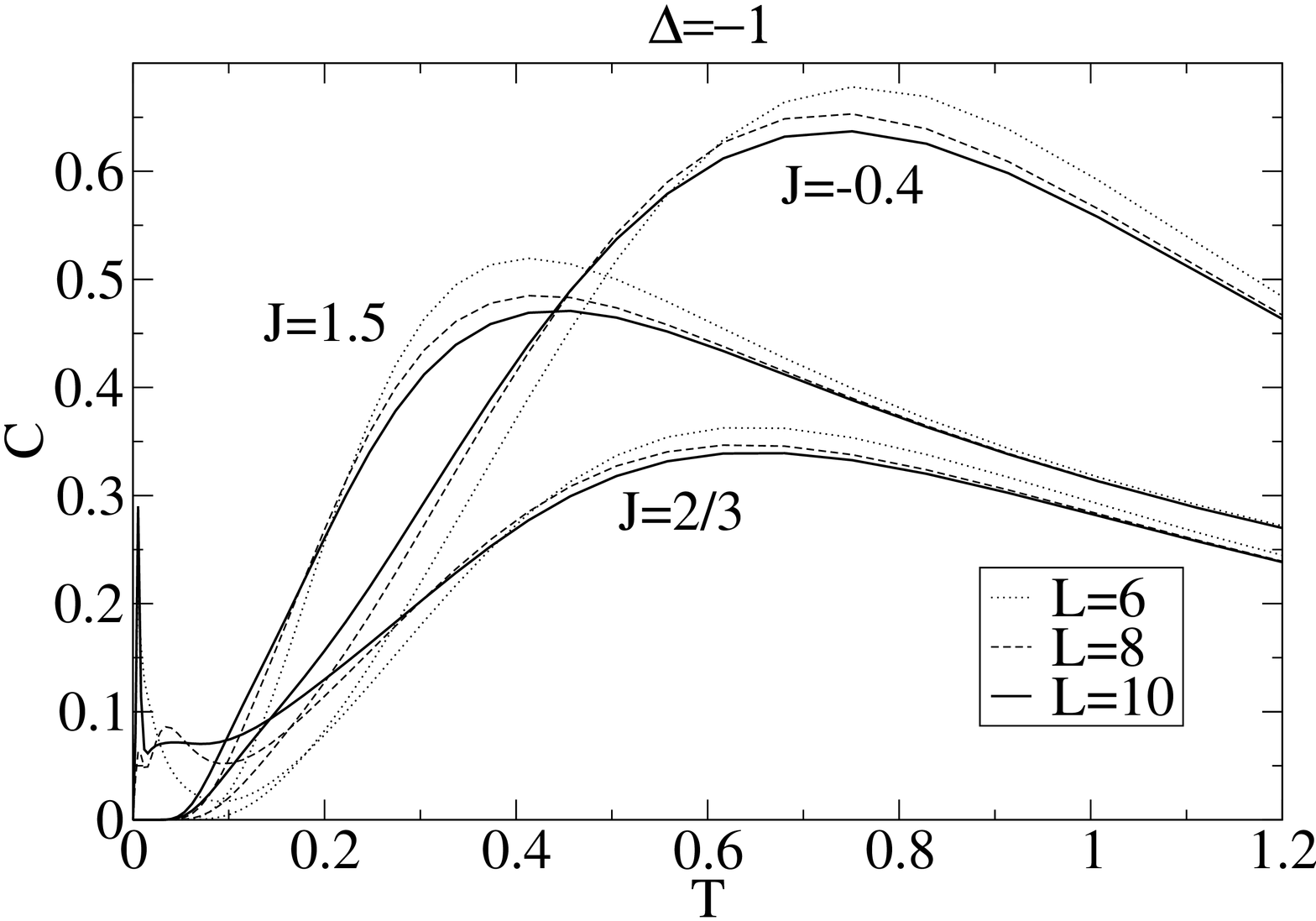}
\includegraphics[width=5.5cm,height=7.9cm,angle=-90]{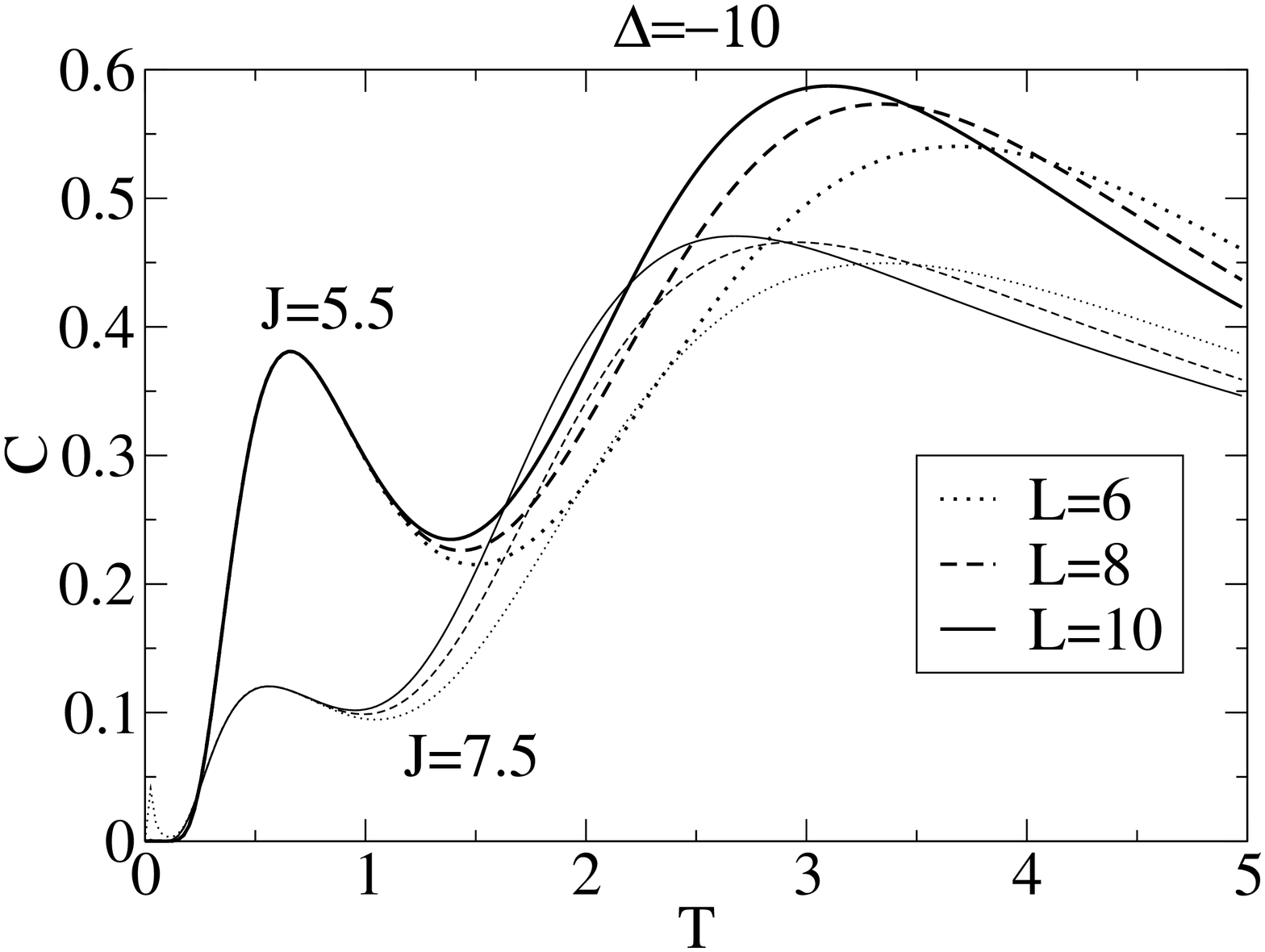}
\caption[fig2]{The behavior of the specific heat for the finite rings $L=6,8,10$  in the ferrimagnetic ($\Delta=-1$) region and in the massive antiferromagnetic 
region ($\Delta=-10$).}
\label{fig_Csp_n}
\end{center}
\end{figure}

In order to study the specific heat properties at higher temperatures we 
diagonalize numerically the full eigenspectrum of the quantum chain 
\eqref{hamiltonian} with $L=6,8$ and $10$, and compute their specific heat. 
These results are shown in Fig. \ref{fig_Csp_n} for the ferrimagnetic models 
$\Delta =-1$, with the values of $J=\frac{3}{2},\frac{2}{3}$ and
$-\frac{2}{5}$,  
and for 
$\Delta =-10$ for $J=5.5$ and $7.5$. These curves except at low temperatures 
give us a good idea about their convergence to the bulk limit 
$L\rightarrow \infty$. 
At low temperatures these  curves display some anomalous behavior, with small 
peaks and inflexion points. These anomalous behavior are due to the 
combination of the effect of the zero temperature phase transitions, that is not 
quite sensitive to the lattice size, with the normal finite-size 
effects, as normally happens when we study a finite chain  that 
becomes massless as $L\rightarrow \infty$. These effects can be 
distinguished  (see Fig. \ref{fig_Csp_n}) as the lattice size grows 
(see also Fig \ref{Low_T_Csp}). 
We also show in Fig. \ref{fig_Csp_n}  some $C_L(T,\Delta,J)$ for the ferrimagnetic 
point $\Delta =-10$. In this case as we can see in Fig. \ref{fig_DJ}
we have approximately a direct first order phase transition happening at 
$J=J_0(\infty,\Delta) = -\frac{2}{3}\Delta=\frac{20}{3}$. 
Consequently for $J=J_0(\infty,\Delta)$ we should
observe the anomalous finite-size behavior as mentioned before (see 
the small peak at $J \sim 5.5$ and $T\sim 0.02$), that disappears as $J$ 
increases. In the vicinity of $J=J_0(\infty,\Delta)$ we observe a small peak which is a signature of ferrimagnetism. As we departure from the 
phase transition region $J \gg J_0(\infty,\Delta)$ this small peak disappears and we 
have only the standard Schottky-like peak of the specific heat.

\begin{figure}[ht]
\begin{center}
 \psfrag{C}[b][][0.9]{$C_L(T,\Delta,J)/\kB$}
\psfrag{J}[][]{\small$J$}
\psfrag{D=-1}[][]{$\Delta=-1$}
\psfrag{D=-2}[][]{$\Delta=-2$}
\includegraphics[width=5.5cm,height=7.7cm,angle=-90]{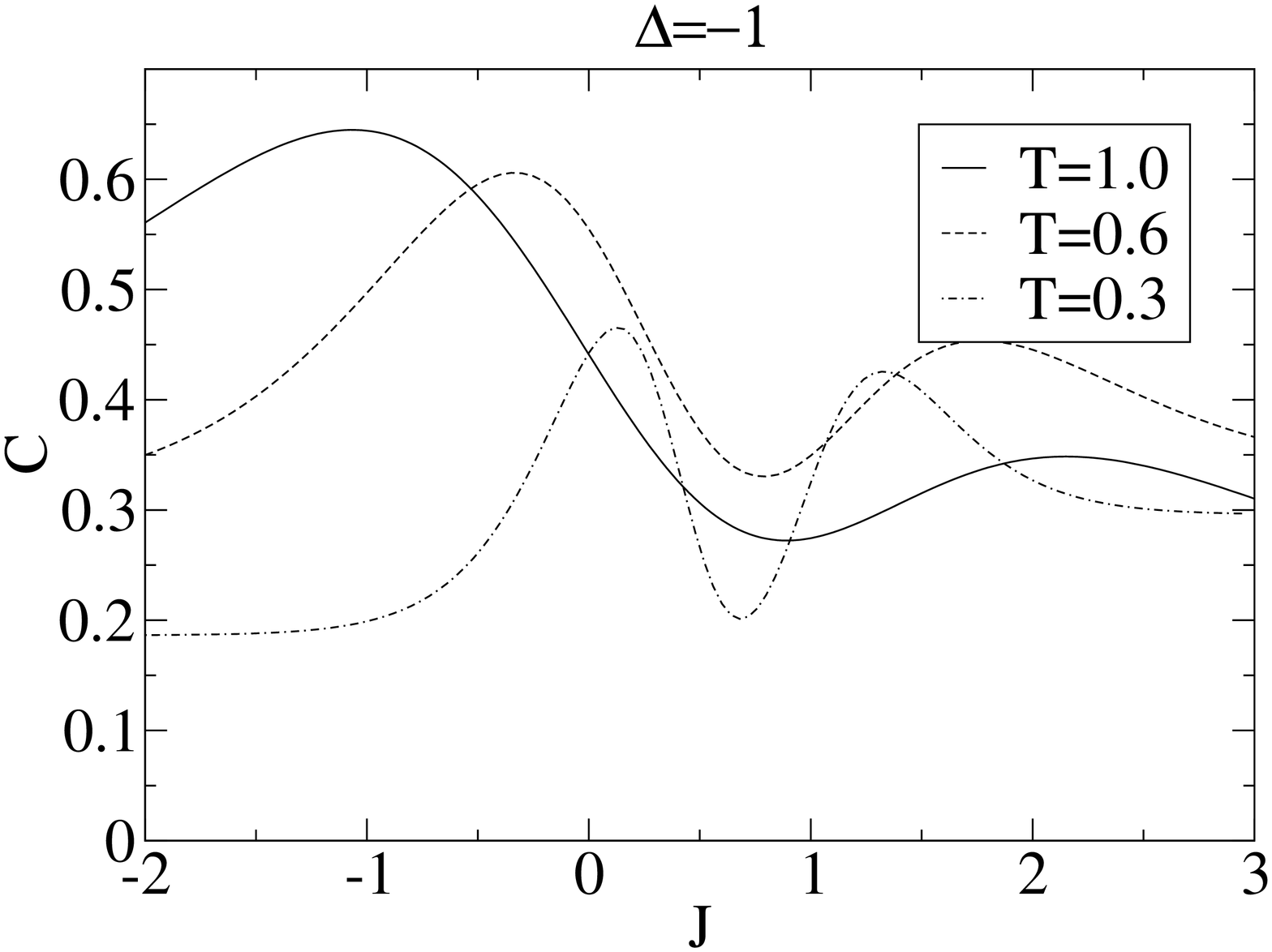}
\includegraphics[width=5.5cm,height=7.7cm,angle=-90]{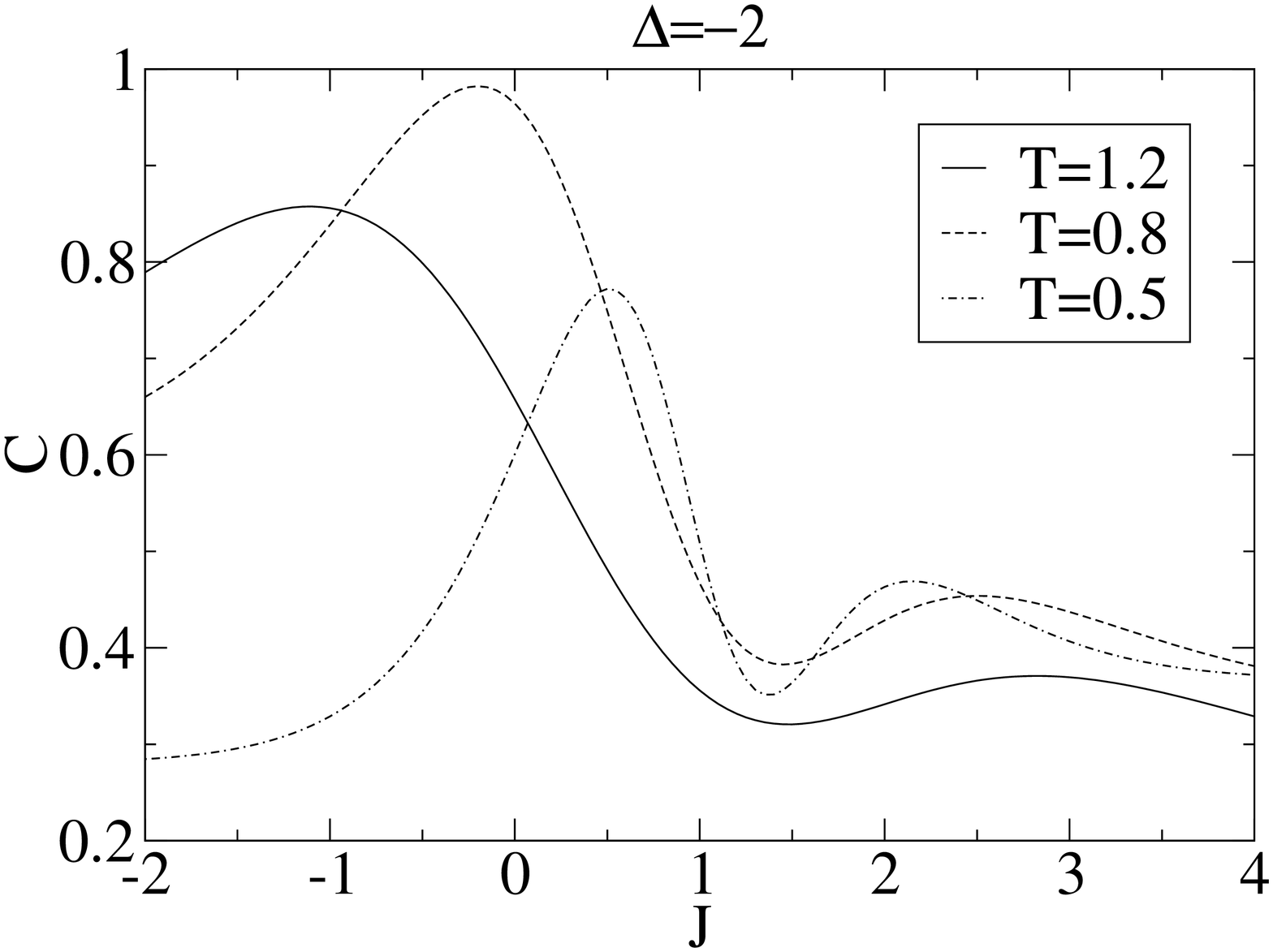}
\caption[fig2]{Specific heat as a function of $J$ for  fixed temperatures in the ferrimagnetic ($\Delta=-1$)  and antiferromagnetic ($\Delta=-2$)  regions.}
\label{fig_Csp1}
\end{center}
\end{figure}

From now on we restrict our calculations with the finite ring with 
$L=10$ sites. As illustration we plot in Fig. \ref{fig_Csp1}  the specific 
heat as a function of $J$ for some temperatures and for the anisotropies 
$\Delta = -1$ and $\Delta =-2$. For other values of $\Delta$ the general 
feature of two maximums repeats. The reason for the occurrence of these 
two maximums can be understood from Fig. \ref{Low_T_Csp}.  The sharp 
peak that happens in Fig. \ref{Low_T_Csp} for low temperatures is 
smoothed as the temperature increases. Similarly as in Fig.  
\ref{Low_T_Csp} the minimum between the two peaks happens in the region 
where we have the series of phase transitions for the model at 
zero temperature, i. e., $0.6667 \leq J \leq 0.8063$ for 
$\Delta =-1$ and $1.333 \leq J \leq 1.369$ for $\Delta =-2$ 
(see Fig. \ref{fig_DJ}). The maximums appearing for 
$J<J_c$ and $J>J_c$  are due to the eigenlevels of the spin-$\frac{1}{2} $
XXZ and spin-($\frac{1}{2},\frac{3}{2}$) Heisenberg chain, respectively. 
Both maximums diminish as $|J| \rightarrow \infty$ recovering the 
standard specific heat of the spin-$\frac{1}{2}$ Heisenberg chain 
($J\rightarrow \infty$) or spin-($\frac{1}{2},\frac{3}{2}$) Heisenberg chain
($J\rightarrow -\infty$). 

\begin{figure}[ht]
\begin{center}
 \psfrag{C}[b][][0.9]{$C_L(T,\Delta,J)/\kB$}
 \psfrag{T}[]{\small$\kB T$}
\psfrag{J}{$J$}
\psfrag{D=-1}[][]{$\Delta=-1$}
\psfrag{D=1}[][]{$\Delta=1$}
\includegraphics[width=5.5cm,height=7.7cm,angle=-90]{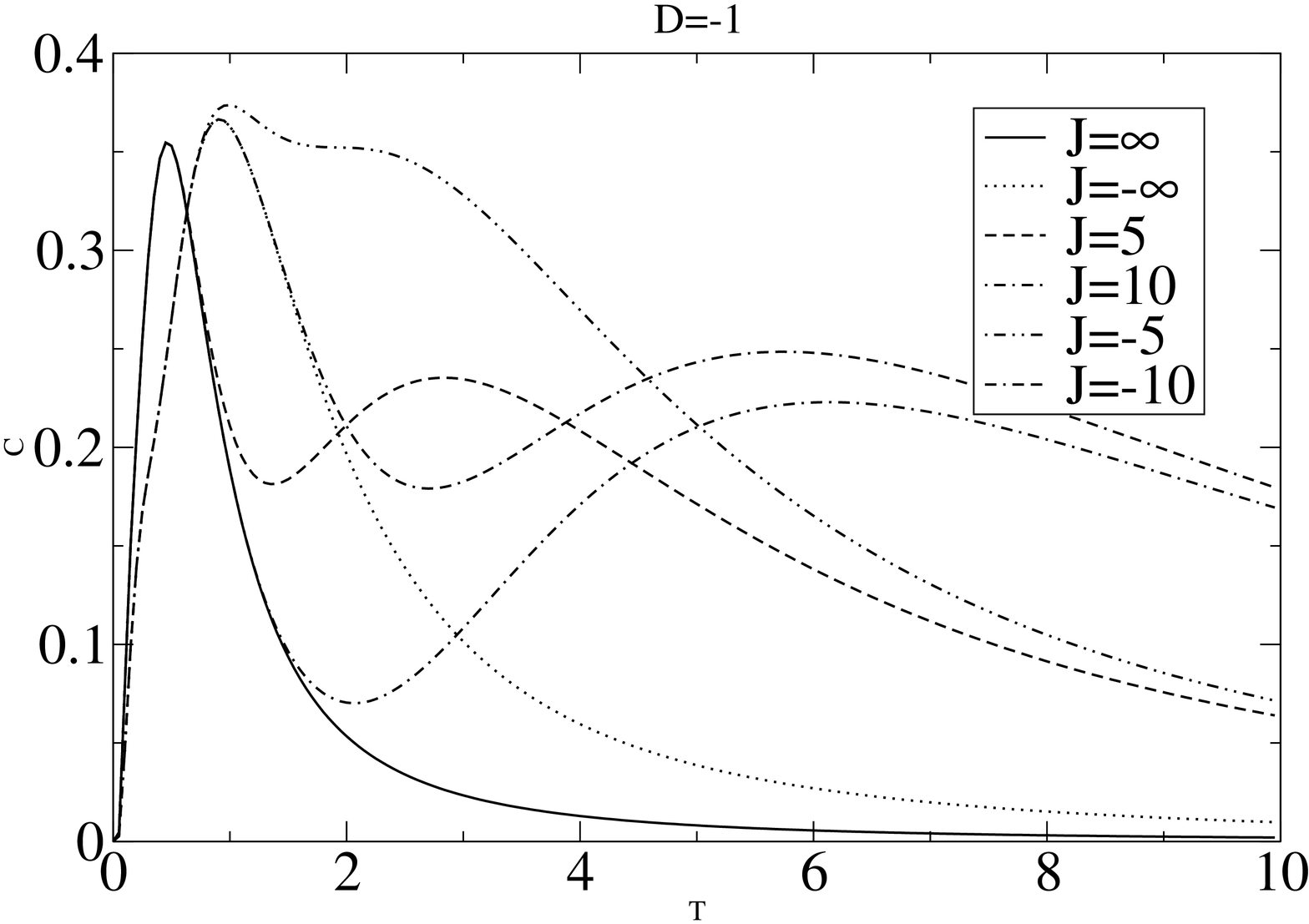}
\includegraphics[width=5.5cm,height=7.7cm,angle=-90]{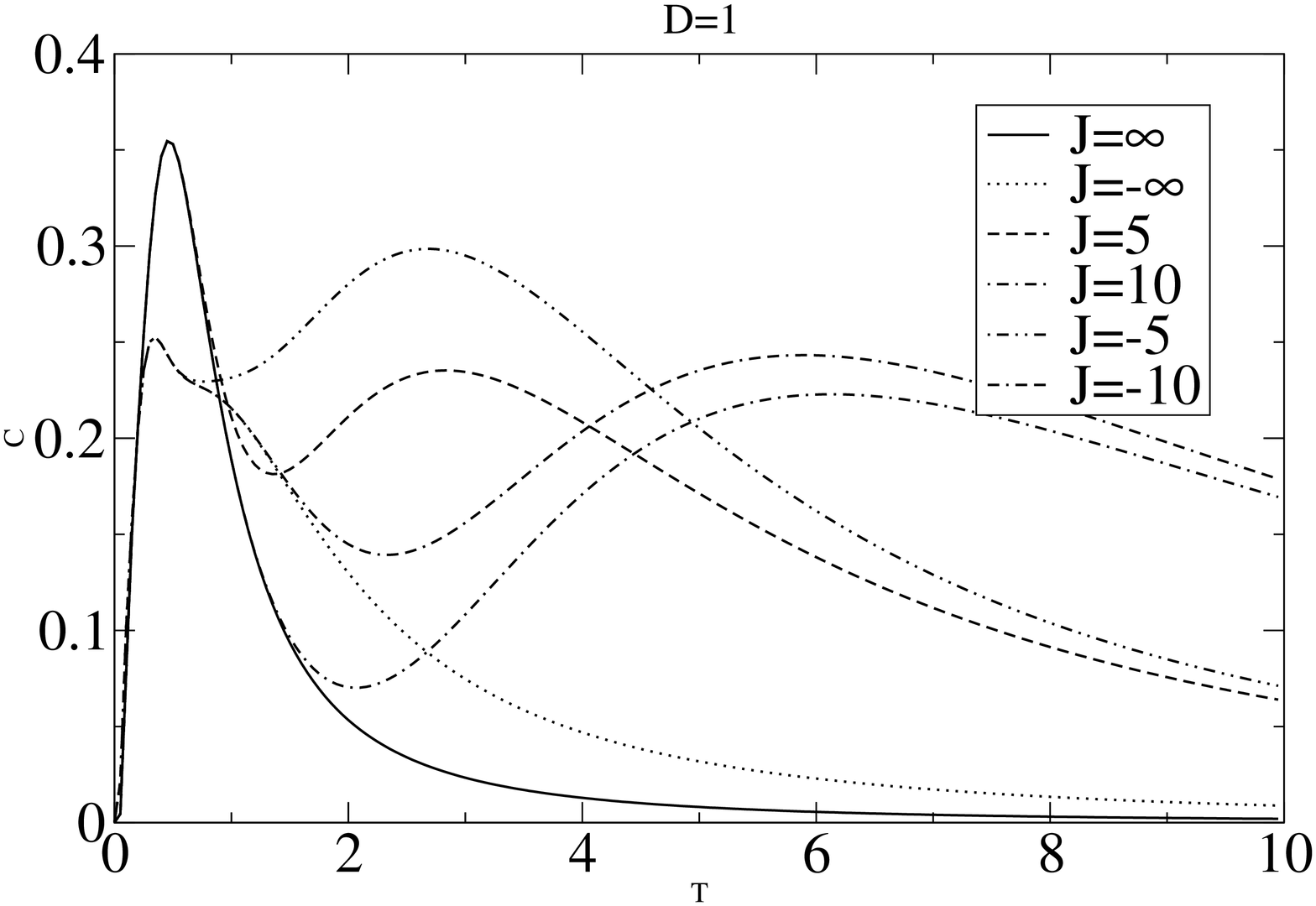}
\caption[fig2]{Specific heat as a function of $T$ for fixed values of the parameter  $J$ in the ferrimagnetic ($\Delta=-1$) and in the ferromagnetic ($\Delta=1$) cases.}
\label{fig_Csp2}
\end{center}
\end{figure}

In order to give an overview about the specific heat in both regions of 
ferromagnetic and ferrimagnetic behavior we plot in Fig. \ref{fig_Csp2} 
the specific heat as a function  of the temperature for several values 
of $J$ and for the anisotropy values $\Delta =-1$ and $\Delta =1$. At the limits $J\rightarrow 
\infty$ and $J \rightarrow -\infty$ the specific heat coincides with that 
of the spin-$\frac{1}{2}$ Heisenberg chain and mixed spin-($\frac{1}{2},
\frac{3}{2}$) Heisenberg chain, respectively. 
Let us consider the ferrimagnetic case $\Delta =-1$. For $J=10$ 
($J=-10$) we see for $\kB T \lesssim 1.5$ the specific heat behavior 
of the  spin-$\frac{1}{2}$ Heisenberg chain (spin-($\frac{1}{2},
\frac{3}{2}$) Heisenberg chain), whereas for $\kB T >1.5$ it appears a peak with 
maximum $\kB T \sim 6$. For $J=5$ ($J=-5$) and  $\kB T \lesssim 0.5$ the 
behavior is that of a spin-$\frac{1}{2}$ Heisenberg chain 
(spin-($\frac{1}{2},\frac{3}{2}$) Heisenberg chain), whereas for 
$\kB T >0.5$ the curves deviate from the $J\rightarrow \infty$ 
($J\rightarrow -\infty$) cases to form another peak with maximum at 
$\kB T \sim 2.3$. As we can see in Fig. \ref{fig_Csp2} a similar behavior
appears for the ferromagnetic cases where $\Delta =1$. The second maximum
appearing in Fig. \ref{fig_Csp2} can be understood from the limiting 
case of the model where $|J|\gg\Delta$. In this case we should consider only the
interactions of the spins around the triangles of Fig. \ref{fig1} 
($\tau_i^a,\tau_i^b,\tau_i^c$) and we have a set of $\frac{L}{2}$ decoupled
3-sites chain whose specific heat is given by 
\begin{align}\label{H_T_Spc}
C_{\infty}(T,\Delta,J)=2\kB \Big(\frac{3J}{4\kB T}\Big)^2\Big({\rm e}^{\frac{3J}
{4\kB T}}+{\rm e}^{\frac{-3J}{4\kB  T}}\Big)^{-2}.
\end{align}
This relation gives a good approximation for the second maximum appearing for
positive and negative values of $J$. If we compare this last relation with
\eqref{d5}, which is valid for the ferromagnetic model at low temperatures we
verify that the factor 2 appearing in \eqref{d5} that was responsible for
the presence of two distinct maximums, now in \eqref{H_T_Spc} is missing since
the degeneracy $2^{\frac{L}{2}}$ gives for the decoupled system an irrelevant
contribution.

\begin{figure}[ht]
\begin{center}
 \psfrag{C}[b][][0.9]{$C_L(T,\Delta,J)/\kB$}
 \psfrag{T}[]{\small$\kB T$}
\psfrag{J=0.75}[][]{$J=0.75$ and $L=10$}
\psfrag{D=-1.3}[][][0.8]{$\Delta=-1.3$}
\psfrag{D=-0.5}[][][0.8]{$\Delta=-0.5$}
\psfrag{D=0.5}[][][0.8]{$\Delta=0.5$}
\psfrag{D=1}[][][0.8]{$\Delta=1$}
\includegraphics[width=5.5cm,height=7.7cm,angle=-90]{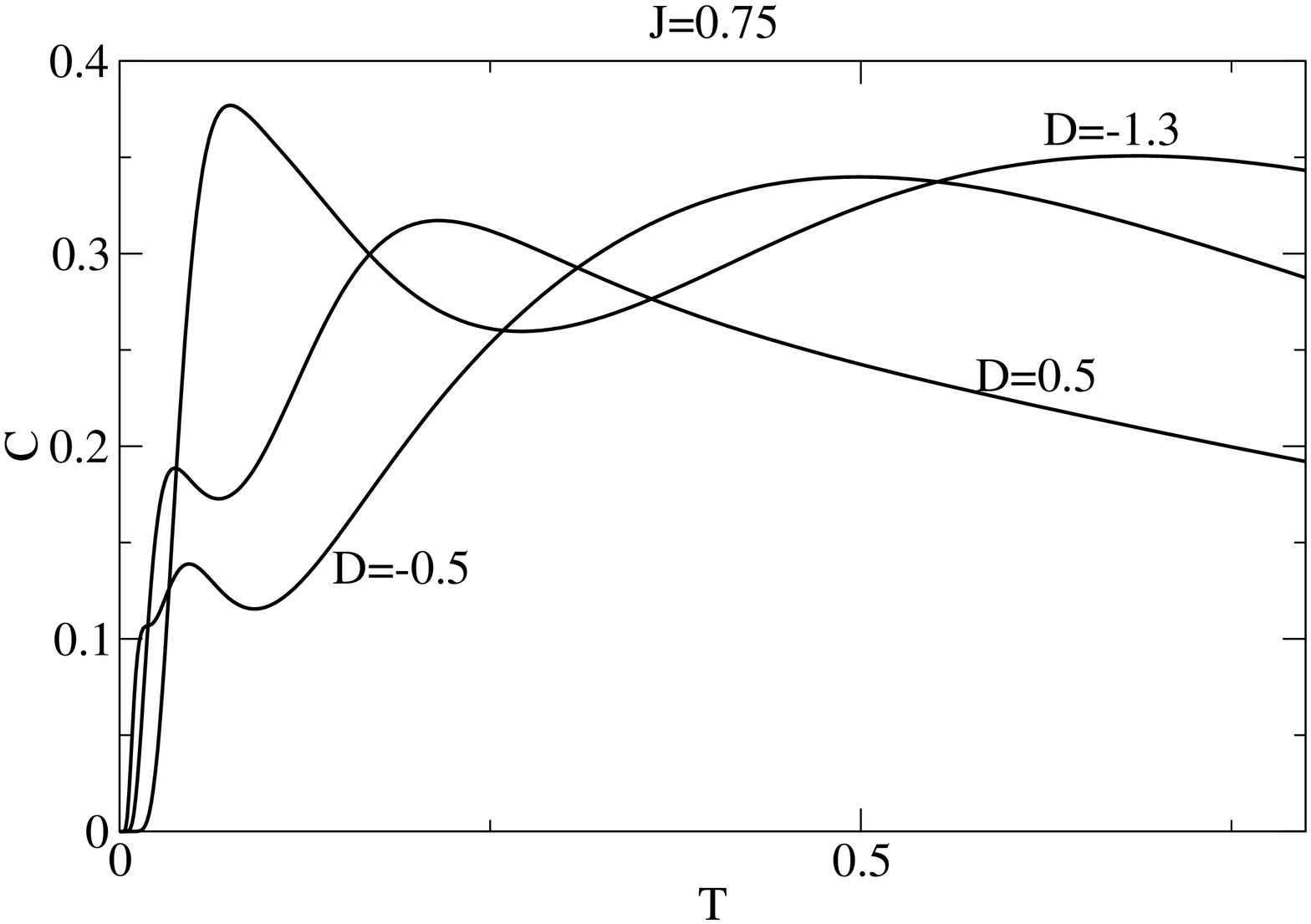}
\caption[fig2]{Specific heat as a function of $T$ for a fixed parameter $J=0.75$ in the massive antiferromagnetic ($\Delta=-1$), massles antiferromagnetic ($\Delta=-0.5$) and massles ferromagnetic ($\Delta=0.5$) cases.}
\label{fig_Csp_D}
\end{center}
\end{figure}

In Fig. \ref{fig_Csp_D} we illustrate the distinct behavior of the specific heat, as a function of temperature, in the several phases of the model. The curves are for the $L=10$ chain and are plotted for a fixed value of the parameter $J=0.75$ and for some values of the anisotropy $\Delta$.  All curves show beyond the standard high temperature broad peak another one at low temperatures due to the proximity of the zero temperature phase transitions. The hight of these peaks depend on their location relate to the nearby zero temperature phase transition  points (see Fig.\ref{fig_DJ} and Fig.\ref{Low_T_Csp}).

\subsection{The magnetic susceptibility}

In order to  calculate the magnetic susceptibility 
of our tetrahedral spin chain \eqref{hamiltonian} we introduce in
\eqref{hamiltonian} the additional term
\begin{align}\label{20}
{\bf B}=-g\uB h\sum_{i=1}^{L/2}\big(\sigma^z_i+\tau^z_i),
\end{align}
where $h$ is an  external magnetic field,   $g$ is the Land\'e factor and $\uB$ is the Born magneton.
This is equivalent to the addition of the term 
\begin{align}\label{20p}
{\bf B}=-g\uB h\sum_{i=1}^{L/2}\big(\sigma^z_{2i-1}+S^z_{2i})
\end{align}
in the equivalent mixed spin Hamiltonian \eqref{Hamt_g}.
\begin{widetext}
We calculate in this section the zero-field magnetic susceptibility per site
$\chi_L(T,\Delta,J)$ for our tetrahedral spin chain, i. e., 
\begin{align} \label{20pp} 
\chi_L(T,\Delta,J) = \frac{\kB T}{L} \frac{\partial^2}{\partial h^2}\ln
{\cal Z}_L(T,\Delta,J,h)|_{h=0}, 
\end{align}
where ${\cal Z}_L(T,\Delta,J,h)$ is the partition function \eqref{d0} with the
inclusion of the magnetic interactions \eqref{20} or \eqref{20p}. 

Let us consider initially the low and high temperature limits for the magnetic
susceptibility. At low temperatures the degeneracy $d(J,\Delta)$ of the ground
state, that depends on the value of $J$ and $\Delta$ (see the phase diagram in
Fig. \ref{fig_DJ}) give us the behavior

\begin{align}\label{lowTX}
\lim_{T\rightarrow 0}T\chi_{_L}\!(T,\Delta,J)=\frac{g^2\uB^2}{\kB}\frac{m(m+1)}{3L}, \quad 
d(J,\Delta) = 2m+1.
\end{align}
\end{widetext}
On the other hand at high temperatures $\chi_L(T,\Delta,J)$ should be
independent of the field strength and saturate the Curie law value for
independent spins. The magnitude of the independent spins depend on the value
of the parameter $J$. In the cases where $|J| \gg\Delta$ we can neglect the
anisotropy $\Delta$ in \eqref{Hamt_g} and we have the asymptotic results

\begin{align}\label{highTX}
T\chi_L(T,J) = \frac{g^2\mu_b^2}{4\kB T} ( 
1 + \frac{2e^{-\frac{3J}{4\kB T}}}
{e^{-\frac{3J}{4\kB T}} + e^{\frac{3J}{4\kB T}}}  ).
\end{align}
This means that when $J\rightarrow -\infty$ we recover the Curie law for the 
spin-($\frac{1}{2},\frac{3}{2}$) Heisenberg mixed chain: 
$T\chi_L(T,\Delta,J) = \frac{3}{4}\frac{g^2\mu_B^2}{\kB}$, while for $J\rightarrow 
\infty$ we obtain the Curie law for the spin-$\frac{1}{2}$ XXZ chain: 
$T\chi_L(T,\Delta,J) = \frac{1}{4}\frac{g^2\mu_B^2}{\kB}$. At finite values of 
$|J|$, but $T \rightarrow \infty$, we obtain from \eqref{highTX}
a third possibility: 
$T\chi_L(T,\Delta,J) = \frac{1}{2}\frac{g^2\mu_B^2}{\kB}$. 
We remark that the values of the constants in these Curie laws are 
independent on the  existence of an infinite number of phase transitions or
a single one, as it happens in the ferromagnetic regime.

\begin{figure}[ht]
\begin{center}
 \psfrag{TX}[b][][0.9]{$T\chi_{_L}(T,\Delta,J)$}
 \psfrag{T}[]{$T$}
 \psfrag{J=inf}[]{$J=\infty$}
\includegraphics[width=6cm,height=7.7cm,angle=-90]{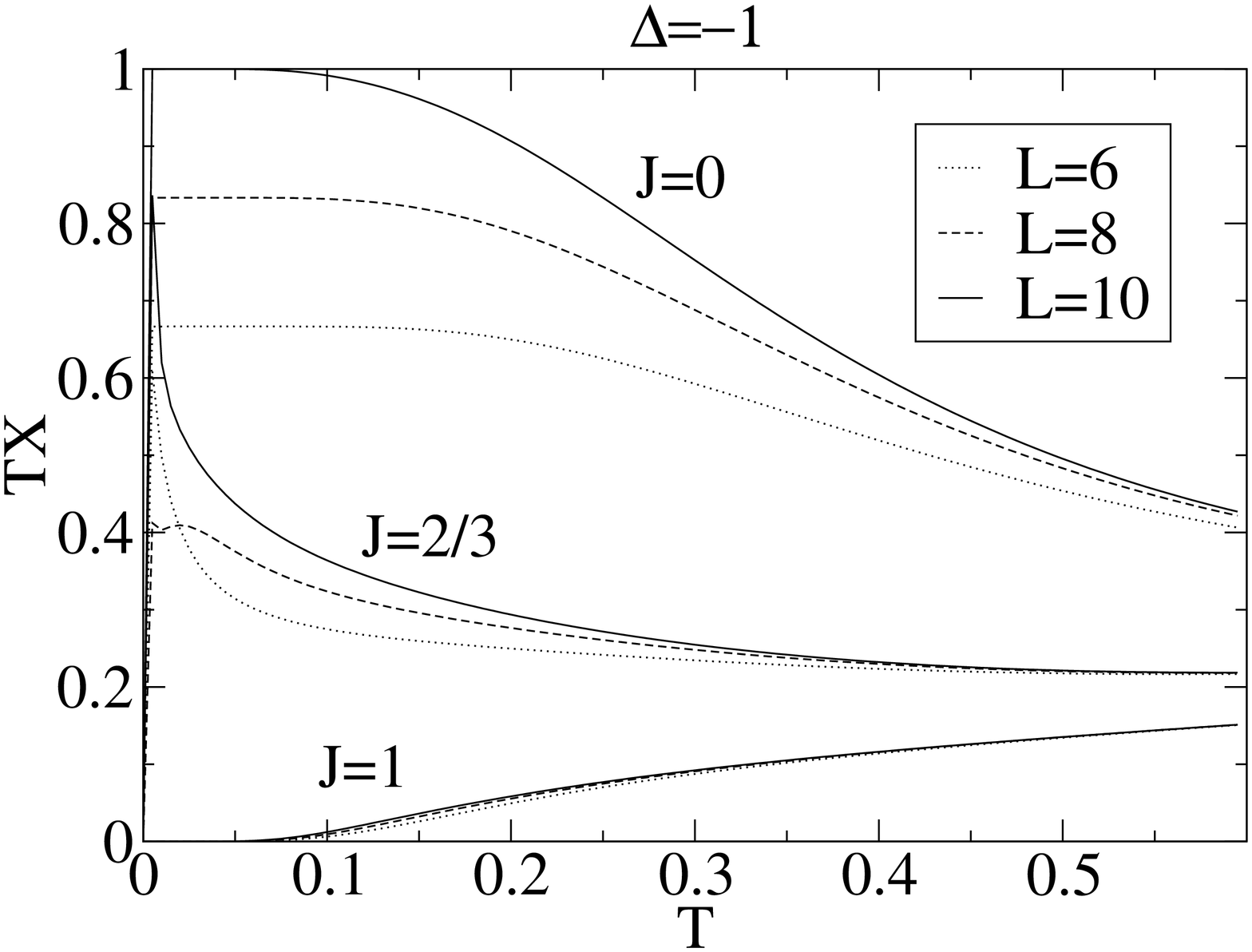}
\caption[fig2]{The magnetic susceptibility  as a function of the temperature for finite rings in the ferrimagnetic ($\Delta=-1$) region.}
\label{fig_Xsm_n}
\end{center}
\end{figure}

We calculated numerically the magnetic properties of our system through  
a direct diagonalization of the Hamiltonian for finite chains up to 
$L= 10$ sites. In Fig. \ref{fig_Xsm_n} we show 
$T\chi_L(T,\Delta,J)$ as a function of the temperature for the finite 
chains with $L=6,8$ and $10$ sites in the ferrimagnetic case 
($\Delta =-1$) and the values of $J= 1,\frac{2}{3}$ and $0$. At $J=1$ the 
product $T\chi_L(T,\Delta,J) \rightarrow 0$ when $T \rightarrow 0$. In this
limit it behaves as the spin-$\frac{1}{2}$ Heisenberg chain whose 
ground-state energy is non-degenerated ($m =0$), in agreement with the 
relation \eqref{lowTX}. On the other hand at $J= 2/3$ we have a first order 
phase transition at $T=0$ connecting the phases with similar behavior as the 
spin-($\frac{1}{2},\frac{1}{2},\frac{1}{2},\frac{3}{2}$) 
and spin-($\frac{1}{2},\frac{3}{2}$) Heisenberg mixed chain. 
 Finally at $J=0$ we are in a regime with similar behavior as the mixed spin-($\frac{1}{2},\frac{3}{2}$) Heisenberg chain with a ground-state 
degeneracy $d(\Delta,J) = d(-1,0) = 2\frac{L}{2}+1$. In this case, at 
low temperatures, we obtain the limiting value 
$T\chi_L(T,\Delta,J) \rightarrow \frac{L+2}{12} \frac{g^2\mu_B^2}{\kB}$ predicted by 
\eqref{lowTX}, that in the bulk limit goes to infinity. 

\begin{figure}[ht]
\begin{center}
 \psfrag{TX}[b][][0.85]{$\kB T\chi_{_L}\!(T,\Delta,J)/(g^2\uB^2)$}
 \psfrag{T}[]{$\kB T$}
\psfrag{D=-1}[][]{$\Delta=-1$}
\psfrag{D=1}[][]{$\Delta=1$}
\includegraphics[width=5cm,height=7.7cm,angle=-90]{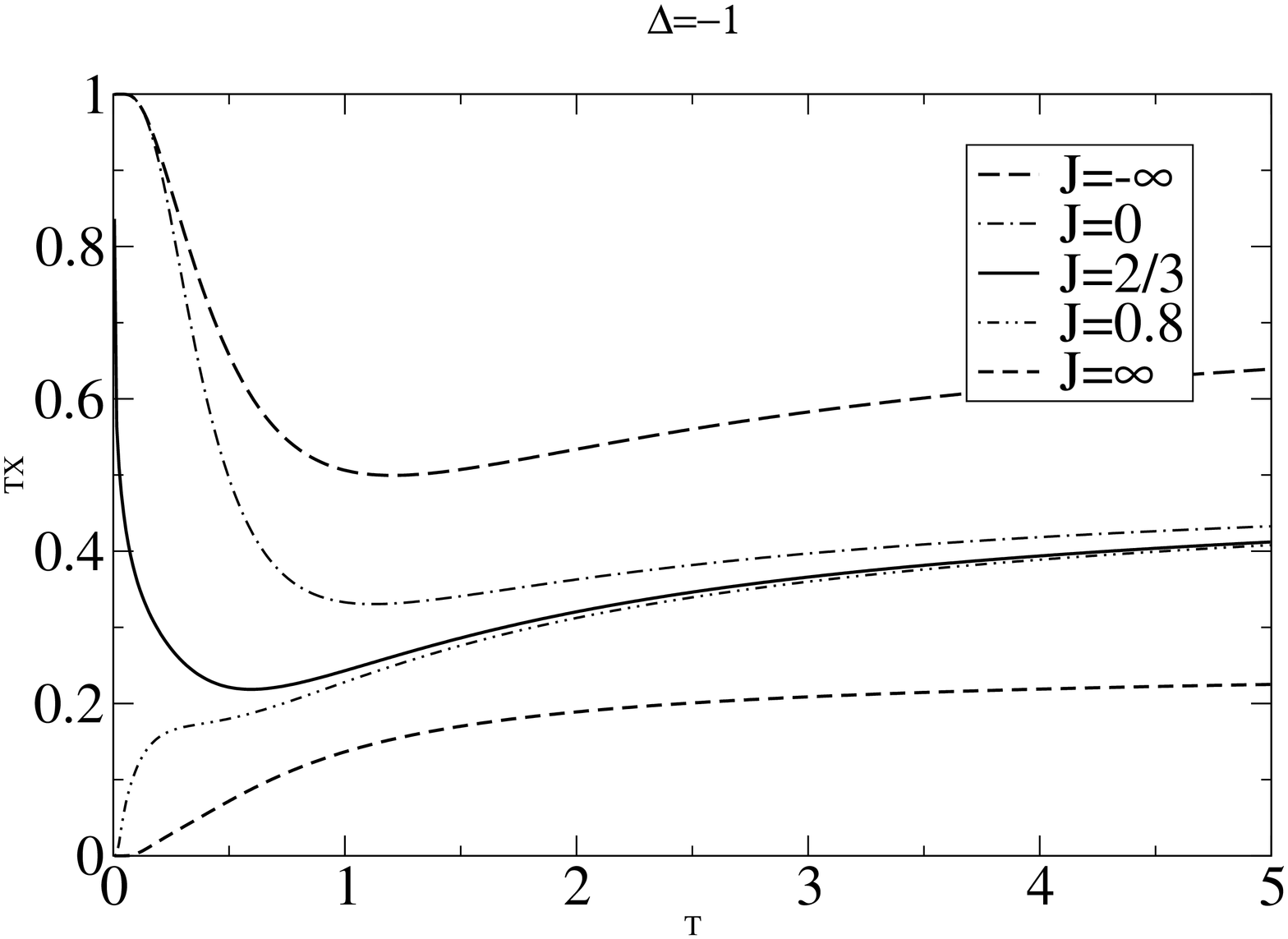}
\includegraphics[width=5cm,height=7.7cm,angle=-90]{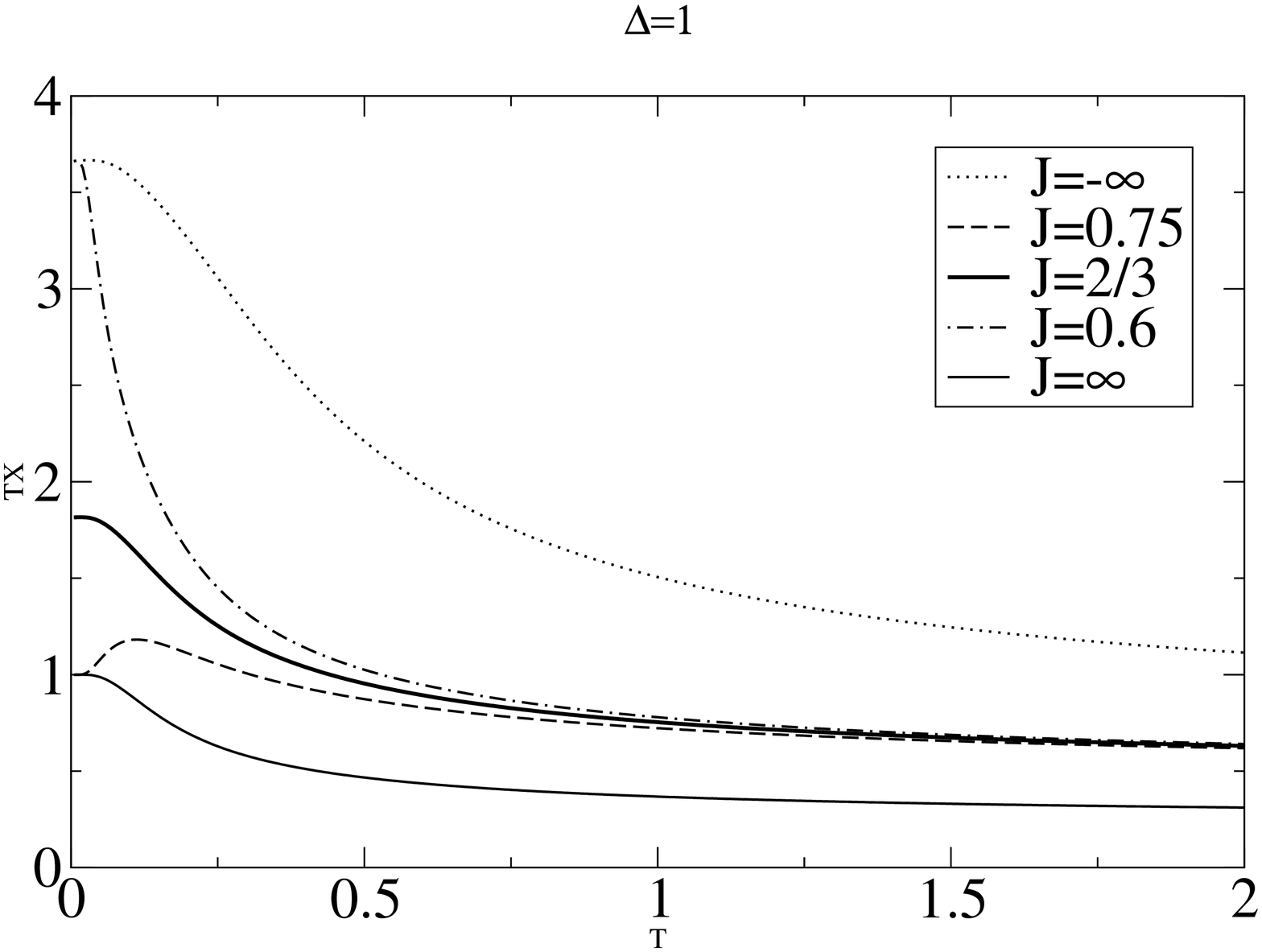}
\caption[fig2]{The magnetic susceptibility  as a function of the temperature for fixed values of $J$, in the ferrimagnetic ($\Delta=-1$) and ferromagnetic ($\Delta=1$) regions.}
\label{fig_Xsm}
\end{center}
\end{figure}

From now on we discuss the magnetic properties only for the finite-size 
chain with $L=10$ sites. In Fig. \ref{fig_Xsm} we plot 
$T\chi_L(T,\Delta,J)$ as a function of the temperature for some values 
of $J$, in the ferrimagnetic case ($\Delta =-1$) and in the 
ferromagnetic case ($\Delta =1$) with zero external magnetic field. 
Let us consider initially the ferrimagnetic case ($\Delta = -1$).  
The limiting case $J \rightarrow \infty$ (long-dashed line) give us the 
spin-($\frac{1}{2},\frac{3}{2}$) Heisenberg chain with a ground-state 
degeneracy $d(\Delta,J) = d(-1,-\infty) = 2\frac{10}{2} +1$. As the
temperatures increases the magnetic susceptibility decreases with a minimum 
value at $\kB T \sim 1.0$ 
 and then increases towards
the value $\frac{3}{4}\frac{g^2\uB^2}{\kB^2}$, in agreement with
\eqref{highTX}. At $J=0$ (doted-dashed line) we have a similar behavior as the
spin-($\frac{1}{2},\frac{3}{2}$) Heisenberg chain 
at very low temperatures whereas for high temperatures the system behaves, 
in agreement with \eqref{highTX}, as $T\chi_L(T,\Delta,J) \rightarrow \frac{1}{2} 
\frac{g^2\uB^2}{\kB}$ (finite $J$).
We also show in this figure the curve with  $J=\frac{2}{3}$ 
(solid line),  i. e., the 
 zero temperature transition point connecting the phases  
 $J_{\frac{1}{2}}$ and $J_{\frac{1}{4}}$. 
 For  $J= 0.8$ (doted-doted-dashed line) we see in the figure that the low 
 temperature behavior tends toward the behavior of the 
 $J \rightarrow -\infty$ case, i. e., 
  the standard behavior of the XXZ chain, where $T\chi_L(T,\Delta,J) \rightarrow 0$ in agreement
  with \eqref{lowTX}, since the degeneracy of the ground state is now finite. 

  In order to compare with the ferromagnetic chains we also show these curves
  for the ferromagnetic case. The degeneracy at $T=0$ is now large (infinite for 
  $L\rightarrow \infty$), but similarly as in the $\Delta =-1$ case depends on
  the value of $\Delta$ (see the phase diagram of Fig. \ref{fig_DJ}).  The magnetic susceptibility in the ferromagnetic regime decreases monotonically at high temperatures tending towards the Curie law behavior with constants given by \eqref{highTX}.

\section{Conclusions}

The phase diagram of the coupled tetrahedral Heisenberg chain given by the 
Hamiltonian \eqref{hamiltonian} is obtained. Exploring the local gauge 
symmetry of the model we show that its eigenenergies is given by the 
addition of the eigenenergies of the XXZ chain with an arbitrary 
distribution of spin-$\frac{3}{2}$ impurities at the even sites of the 
lattice. This imply that part of its eigenspectra is just given by 
the eigenenergies of the exactly integrable  XXZ chain. Moreover, in some 
phases the ground 
state energy as well the low lying energies coincides with  the
eigenlevels of the standard XXZ chain, and we have an example of a
non-integrable model where the ground-state, as well the low-lying 
eigenenergies  are 
exactly known (there  are an infinite number of them 
in the thermodynamic limit). 
 
 The pure XXZ chain exhibits a massless disordered phase for 
 $-1 \leq \Delta \leq 1$ and massive phases ferromagnetically 
 ($\Delta >1$) or antiferromagnetically ($\Delta <-1$) ordered.  The addition 
 of spin-$\frac{3}{2}$ impurities on the XXZ chain brings the possibility of
 ferrimagnetic order on the system due to the Lieb-Mattis theorem
 \cite{liebmattis}. As a consequence the phase diagram of the Hamiltonian 
 \eqref{hamiltonian} (see Fig. \ref{fig_DJ}) is  quite rich, exhibiting  an 
 infinite number of phase transitions connecting phases with all the above
 types of long range magnetic order. In the intermediate phases the ferrimagnetic order 
 increases until the saturated antiferromagnetic order is reached.  It is 
 interesting to note that in the phases $A_{\frac{1}{2},\frac{1}{2}}$, 
 $D_{\frac{1}{2},\frac{1},{2}}$, $F_{\frac{1}{2},\frac{1}{2}}$ as well
 $D_{\frac{1}{2},\eta}$ (see Fig. \ref{fig_DJ}) the degeneracy of the ground 
 state is extensive with the lattice size. These degenerated eigenenergies 
 are connected even at zero temperature through quantum fluctuations and 
 we should expect consequently a finite residual entropy per site  at 
 zero temperature, similarly as happens with the ice \cite{pauling}. 

 The thermodynamical properties of our tetrahedral spin chain was also 
 studied at low temperatures. Thanks to the local gauge symmetry of the 
 model the partition function at an arbitrary temperature was obtained 
 exactly for chains with up to 10 sites. The specific heat as a function of 
 the temperature in general exhibits  besides an  standard Schottky-like peak 
 an additional peak, due to the additional eigenspectrum contribution 
 to the density of states coming from the distinct gauge sectors. Close 
	to the quantum phase transition couplings these peaks fuse into a single
	one. The magnetic susceptibility at zero magnetic field and low 
	temperatures also presents some interesting features. As we change the
	exchange parameter $J$ it shows a minimum for the antiferromagnetic 
	case and a broad maximum for the ferromagnetic case. Moreover at high
	temperatures, depending on the value of $J$, we obtain the Curie law 
	behavior with three distinct constants.

{\it  Acknowledgments}. This work was supported
in part by the Brazilian agencies FAPESP and CNPQ/CLAF (Brazil).


\begin{thebibliography}{99}
\bibitem{conjec} This conjecture could be tested by larger lattices spectral calculations by using the DMRG technique (see Ref. \cite{white}).
\end{thebibliography}
\end{document}